\documentclass[pra,reprint]{revtex4-1}
\pdfoutput=1
\usepackage[T1]{fontenc}
\usepackage[utf8]{inputenc}
\usepackage{amsmath}
\usepackage{amssymb}
\usepackage{amsthm}
\usepackage{hyperref}
\usepackage[caption=false]{subfig}
\usepackage{mathtools}
\usepackage{physics}
\usepackage{dsfont}
\usepackage{tikz}
\usepackage{placeins}
\usepackage[nameinlink,capitalize]{cleveref}
\allowdisplaybreaks

\begin{document}
\title{Memory-assisted long-distance phase-matching quantum key distribution}
\date{\today}
\author{Frank Schmidt}
\email{fschmi@students.uni-mainz.de}
\affiliation{Institute of Physics, Johannes Gutenberg-Universit\"at Mainz, Staudingerweg 7, 55128 Mainz, Germany}
\author{Peter van Loock}
\email{loock@uni-mainz.de}
\affiliation{Institute of Physics, Johannes Gutenberg-Universit\"at Mainz, Staudingerweg 7, 55128 Mainz, Germany}
\begin{abstract}
 We propose a scheme that generalizes the loss scaling properties of twin-field or phase-matching QKD related to a channel of transmission $\eta_{total}$ from $\sqrt{\eta_{total}}$ to $\sqrt[2n]{\eta_{total}}$ by employing $n-1$ memory stations with spin qubits and $n$ beam-splitter stations including optical detectors. Our scheme's resource states are similar to the coherent-state-based light-matter entangled states of a previous hybrid quantum repeater, but unlike the latter our scheme avoids the necessity of employing $2n-1$ memory stations and writing the transmitted optical states into the matter memory qubits. The full scaling advantage of this memory-assisted phase-matching QKD (MA-PM QKD) is obtainable with threshold detectors in a scenario with only channel loss. We mainly focus on the obtainable secret-key rates per channel use for up to $n=4$ including memory dephasing and for  $n=2$ (i.e. $\sqrt[4]{\eta_{total}}$-MA-PM QKD assisted by a single memory station) for error models including dark counts, memory dephasing and depolarization, and phase mismatch. By combining the twin-field concept of interfering phase-sensitive optical states with that of storing quantum states up to a cutoff memory time, distances well beyond 700 km with rates well above $\eta_{total}$ can be reached for realistic, high-quality quantum memories (up to 1s coherence time) and modest detector efficiencies. Similarly, the standard single-node quantum repeater, scaling as $\sqrt{\eta_{total}}$, can be beaten when approaching perfect detectors and exceeding spin coherence times of 5s; beating ideal twin-field QKD requires 1s. As for further experimental simplifications, our treatment includes the notion of weak nonlinearities for the light-matter states, a discussion on the possibility of replacing the threshold by homodyne detectors, and a comparison between sequential and parallel entanglement distributions.
\end{abstract}
\maketitle

\section{Introduction}

In 1984  Bennett and Brassard presented a protocol (BB84) \cite{BB84} that allows two parties (typically referred to as Alice and Bob) to distribute an information-theoretically secure key exploiting the fundamental laws of quantum mechanics. 
This was the beginning of the new field of quantum key distribution (QKD) leading now to the first commercial applications of quantum technology (see \cite{plobhistory} for a recent overview of QKD). Based on this concept, a key distribution scheme over 421 km of glass fiber was demonstrated recently \cite{PhysRevLett.121.190502}.
Nonetheless, a complication of realistic QKD schemes is the linear scaling of the secret-key rate with the channel transmittance $\eta_{total}$ \cite{TGW}, where $\eta_{total}$ decreases exponentially with the distance, $\eta_{total}=\exp(-L/L_{att})$, where $L_{att}=22$  km is the typical attenuation distance of an optical fiber. In fact, it was shown that this linear scaling for large distances is a fundamental property of point-to-point QKD, expressed by the so-called repeaterless (or `PLOB') bound \cite{PLOB}, $-\log_2\left(1-\eta_{total}\right)$, in terms of secret bits per channel use, where $-\log_2\left(1-\eta_{total}\right)\approx1.44 \eta_{total}$ for $\eta_{total}\ll1$. 

As a consequence, one needs to split the total channel into multiple segments of smaller length in order to overcome the linear scaling. Splitting the total distance into multiple segments of smaller length is the underlying idea of all types of quantum repeaters making use of either quantum memories \cite{repeater_briegel,DLCZ} or quantum error-correcting codes \cite{PhysRevLett.112.250501,PhysRevLett.117.210501,azumarepeater,Munro2012} or both in order to improve the transmission rate.
Due to the quantum mechanical no-cloning theorem it is impossible that a quantum repeater simply reamplifies an incoming optical quantum state at every intermediate station along the channel like for a classical repeater with classical light pulses.
The only experimental demonstration of a quantum repeater so far  overcoming the PLOB bound in terms of a secret key rate per channel use was reported recently in Ref. \cite{memoryassisted_experiment} based on a solid-state light-matter interface and memory system using SiV color centers in diamond. 

Besides its scalability, an essential element of a QKD scheme is its security in a realistic setting.
More than a decade ago it was shown that QKD systems are vulnerable to hacking attacks (see \cite{quantum_hacking,quantum_hacking2} for a review) and it was realized that the typical assumptions of the security proofs are not met in a practical implementation.
Device-independent QKD \cite{DI_Mayer,DI_Vidick} was proposed as a possible solution. Its security proof no longer depends on the actual implementation, since it relies on the violation of a Bell inequality. However, this type of protocol yields only very small secret-key rates.
 A more promising approach in this respect is measurement-device-independent (MDI) QKD \cite{PirandolaMDI,MDI_Lo}, where Alice and Bob send states to a middle station, Charlie, who performs a measurement that can be treated as a black box.
As such, the middle station may be completely untrusted, with Charlie potentially embodied by an eavesdropper, Eve.
 This approach becomes secure against the most problematic class of detector attacks and yields reasonable secret-key rates.

Quite recently it was shown that  MDI QKD, exploiting  interference of phase-sensitive phase-encoded optical states sent from Alice and Bob to Charlie, gives a scaling of the asymptotic secret-key rate of $O(\sqrt{\eta_{total}})$ \cite{twinfield_introduction}, originally named as twin-field QKD. 
Many works have now appeared improving or simplifying the security proof and suggesting variations of this protocol \cite{twinfield_luetkenhaus,tf_ref1,tf_ref2,tf_ref3,tf_ref4,tf_ref5,tf_ref6}. 
For the present work, especially relevant is the version referred to as phase-matching QKD \cite{twinfield_luetkenhaus,tf_ref2}.
Therefore, it is possible, in principle, to overcome the PLOB bound \cite{PLOB}  without making use of quantum memories or quantum error-correcting codes.
 There are already first experimental demonstrations of twin-field QKD that claim to have overcome the PLOB bound \cite{tf_experiment,PhysRevLett.123.100505,tf_experiment_canada,Minder2019,tf_experiment_pan}. 

In this work, we introduce a scheme that is an extension of the twin-field/phase-matching protocol to more than two physical segments (i.e., beyond a single middle station) exploiting quantum memories, similar to Ref. \cite{wehner2} and further extending a four-segment variant of Ref. \cite{wehner2}, but with  single-photon-based single-rail (single-mode) qubits replaced by coherent states. 
Our scheme makes use of quantum memories - a kind of memory-assisted extension of phase-matching QKD \cite{twinfield_luetkenhaus,tf_ref2}, and thus is ideally, with  sufficiently good memories and operations, in principle, scalable to long distances.
The scheme shares similarities with a hybrid quantum repeater (HQR) \cite{hybridquantumrepeater} where an optical coherent state subsequently interacts with two spin-based matter quantum memories and entangles these two spin qubits after a suitable measurement of the optical mode. However, in the original HQR, the optical mode travels all the way from one memory station to another before its detection at that station.
In our scheme, crucially, there will be a middle station, half way between the memories, equipped with a beam splitter and detectors. This way we will be able to generalize the loss scaling behavior of twin-field/phase-matching QKD from an effective channel length of $\frac{L}{2}$ to $\frac{L}{2n}$ for $2n$ physical segments and a total physical channel of length L with only $n-1$ memory stations.
We find that compared with the original HQR based on unambiguous state discrimination \cite{hybridquantumrepeater_usd}, the new MA-PM QKD scheme leads to a scaling advantage where in all relevant quantities $\eta$ (transmittance per repeater segment) becomes $\sqrt\eta$.  
While our scheme could be supplemented by additional quantum error correction or detection mechanisms such as entanglement purification \cite{repeater_briegel,hybridquantumrepeater,hybridquantumrepeater_encoding}, here we shall consider the theoretically and especially experimentally simplest intermediate-scale versions without error correction.
\\
The outline of our paper is as follows. In Sec. \ref{sec:background} we will briefly introduce the main ideas of twin-field/phase-matching QKD, the HQR, and possibilities for generating the entangled states needed for our scheme. In  Sec. \ref{sec:quantumrepeater} we will then describe our version of a new type of HQR and discuss its obtainable secret-key rate by employing a BB84 protocol, and focusing on the channel-loss-only case. For the more general and realistic situation, we will briefly mention different error models - including channel loss, memory dephasing, detector dark counts, phase mismatch, and depolarization, referring to the appendix for details.
We will also briefly describe a variant of our scheme based on optical homodyne measurements, similar to the original HQR \cite{hybridquantumrepeater}. 
Then we will explicitly calculate the attainable secret-key rates in Sec. \ref{sec:comparison} for the first-order generalization (i.e., four physical segments, $n=2$) considering a fairly large and representative set of realistic parameters.
Although our main focus is on secret-key rates per channel use, we will also include a discussion on the usefulness of our scheme in terms of the ultimate figure of merit, the secret-key rate per second.
We conclude in Sec. \ref{sec:conclusion} and give more details about the basic concepts, assumptions and calculations in the appendices.

\section{\label{sec:background}Background}
\subsection{Twin-field/Phase-matching QKD}

There are many different variations of twin-field QKD\cite{twinfield_introduction,twinfield_luetkenhaus,tf_ref1,tf_ref2,tf_ref3,tf_ref4,tf_ref5,tf_ref6} and we will stick to the version of  Ref. \cite{twinfield_luetkenhaus}, since their protocol is conceptually easy to understand and it is very similar to the generalized scheme that we will introduce:
\begin{itemize}
\item{Alice and Bob choose randomly and independent from each other with a probability $p_{mode}$ if the current round is used for key generation or for estimating information leakage (test mode).}
\item{If the key-generation mode is chosen, Alice (Bob) generate uniformly distributed random bits $k_A$  ($k_B$) and send coherent states with amplitude $\alpha e^{i \pi k_{A/B}}$ to an untrusted middle station called Charlie (Alice and Bob pre-agreed upon an $\alpha$). 
If the test mode is chosen, they generate coherent states of an amplitude according to some fixed probability distribution and send the optical states to the middle station.}
\item{If Charlie is honest, he applies a balanced beam splitter (BS) to Alice's and Bob's optical modes and employs threshold (on/off) detectors for the BS output modes, announcing the measurement results. }
These steps are repeated until a long data set is obtained. If Alice and Bob use the key-generation mode and exactly one of the two detectors clicks, $k_a$ and $k_b$ are perfectly (assuming no dark counts) (anti-)correlated depending on which of the two detectors clicked.
In our scheme, the level of  security of these (anti-)correlations that manifests itself in the quality of the randomly phase-flipped entangled (effective) density operator shared by Alice and Bob will depend on the channel transmission, the overlap of the coherent states, and the type of detectors (we shall also consider photon-number resolving detectors, PNRDs).
\item{The usual QKD steps of sifting, estimating the error rate and leaked information, error correction and privacy amplification need to be performed.}
\end{itemize}

Note that a pre-agreed complex amplitude $\alpha$ implies that Alice's and Bob's lasers should not differ in their phase. However, it is also unreasonable to assume that the optical path length between Alice and Charlie perfectly coincides with that of Bob and Charlie.
 Therefore, it is necessary to stabilize Alice's and Bob's laser frequencies and also apply phase stabilization techniques because of the phase drift in the fiber of the communication channel.
 This extra experimental complication in a twin-field/phase-matching QKD scheme is somewhat the price to pay for the scaling gain, $\eta_{\text{total}}\rightarrow\sqrt{\eta_{\text{total}}}$.

Since the untrusted Charlie (who could always be Eve) performs the measurements, the protocol is a MDI protocol \cite{PirandolaMDI,MDI_Lo}, meaning that we are immune to attacks upon the detectors, which seems to be the most vulnerable part in a QKD system.
\subsection{Hybrid quantum repeater}

Each segment of  a so-called HQR consists of two quantum memories placed at its ends \cite{hybridquantumrepeater} and connected by an optical channel.
Each quantum memory is represented by a two-level spin system which is initially in the state $\frac{1}{\sqrt{2}}\left(\ket{\uparrow}+\ket{\downarrow}\right)$. We will consider a light-matter interaction between each memory and a single-mode coherent state of light such that 
\begin{equation}
\label{eq:hybridrepstate1}
\hat{U}_{int}\left(\theta\right)\left(\ket{\uparrow}+\ket{\downarrow}\right)\ket{\alpha}=\ket{\uparrow}\ket{\alpha e^{-i \theta}}+\ket{\downarrow}\ket{\alpha e^{i \theta}}\,.
\end{equation}
Thus, the coherent-state light amplitude is phase-rotated conditioned upon the state of the spin.
We call the result of this interaction a hybrid entangled state and there exist different physical phenomena for obtaining this transformation.
An attractive feature here is that we may even consider a fairly weak interaction, $\theta\ll1$.
A few more details about these interactions will be given in the next subsection.

First we let one memory interact with the optical mode, which is then send to the other memory at the next repeater station where we again apply the light-matter interaction. This results, in the absence of channel loss, in the (normalized) state
\begin{equation}
\label{eq:hybridrepstate}
\frac{\left(\ket{\uparrow,\downarrow}+\ket{\downarrow,\uparrow}\right)\ket{\alpha}+\ket{\uparrow,\uparrow}\ket{\alpha e^{-2i \theta}}+\ket{\downarrow,\downarrow}\ket{\alpha e^{2i \theta}}}{2}\,.
\end{equation}

By discriminating the $\pm2\theta$ phase shifts from the zero phase shift, we can project the two memories onto an entangled Bell state $\ket{\uparrow,\downarrow}+\ket{\downarrow,\uparrow}$.
 Such a discrimination can be performed, for example, by using quadrature homodyne measurements.
In the following, let us assume that $\alpha \in \mathbb{R}^+$. Then we could discriminate the phase shifts by performing a measurement of the momentum-quadrature $\hat{p}:=\frac{1}{2i} \left(\hat{a}-\hat{a}^\dagger\right)$, where $\hat{a}$ and $\hat{a}^\dagger$ are bosonic annihilation and creation operators.
 We can then choose a sufficiently small $\Delta_p$ and if the measurement outcome $p\in[-\Delta_p,\Delta_p]$, we say that we successfully identified a zero phase shift.
 However, this is not  an exact projection onto a Bell state and the  fidelity of the state is a function of the measured value $p$ and $\alpha \sin(2\theta)$,i.e., $2\alpha \theta$ for small $\theta$.
 We could improve the fidelity at the expense of the success probability  by choosing a smaller $\Delta_p$ which means that we are discarding many low-quality states. 
Alternatively, we could also set $\Delta_p$ to a fixed value and increase $\alpha \sin(2\theta)$, however, we cannot increase this value arbitrarily much as soon as the photon loss of the fiber channel is included, since a larger value leads to more decoherence due to the loss.
Therefore one has to find a compromise between average fidelity and raw rate.
For small $\theta$ the  probability of success and the fidelity are only dependent on the transmittance $\eta$ in the repeater segment and on $\alpha \theta$.
One may also consider different measurements on the optical mode such as unambiguous state discrimination based on PNRDs or on/off detectors \cite{hybridquantumrepeater_usd}.
While in our work we discuss both types of measurements, discrete photon and continuous homodyne (App. \ref{app:discuss},\ref{app:homodyne}) detections, the former allow to entirely suppress discrimination errors even for small $\alpha \theta$, thus longer repeater segments are possible. Later we will compare our scheme with a HQR based on unambiguous state discrimination  using on/off detectors.

\subsection{Generation of hybrid entangled states}

States of the form $\ket{\uparrow}\ket{\alpha e^{-i \theta}}+\ket{\downarrow}\ket{\alpha e^{i \theta}}$ are also known as Schr\"odinger cat states, because for large amplitudes of the coherent state they serve as an example of entanglement between a microscopic object like an atom and a macroscopic object like a strong optical field, exactly like in Schr\"odinger's famous thought experiment \cite{Schroedinger1935}.
 In order to realize this in the lab, large efforts have been made to generate these states.
 Mostly the entanglement was generated between the internal state of an atom/ion and its motional degree of freedom or with microwave radiation \cite{RevModPhys.85.1083,PhysRevLett.94.153602,Liao_2006}.
A few other experiments with atom-induced phase shifts  were realized for electromagnetic radiation in the optical frequency domain \cite{Kurtsiefer_phaseshift,Leuchs_phaseshift}. 

We will briefly discuss two different approaches for generating these states. A general advantage of the corresponding physical platform, namely cavityQED with atoms and light, is that it allows for room-temperature operations at optical frequencies, as opposed to solid-state-based approaches such as that of Ref. \cite{memoryassisted_experiment} where sufficient cooling is a necessity.  
One possible approach considers the interaction of light (for a coherent state with amplitude $\alpha$) with a two-level atom (Jaynes-Cummings model \cite{knight} of cavityQED) where the light frequency is largely detuned from the atomic resonance frequency.
The effective interaction Hamiltonian is then given by  
\begin{equation}
\hat{H}_{eff}= \hbar \frac{g^2}{\delta}\left(\hat{\sigma}_+\hat{\sigma}_-+\hat{a}^\dagger \hat{a} \hat{\sigma}_z \right)\,,
\end{equation}
in the regime of large detuning $\delta$ (see for example, Ref. \cite{knight}).
Here, $g$ denotes the coupling constant, $\hat{\sigma}_\pm$ are atomic transition operators and $\hat{\sigma}_z$ is the Pauli-$Z$ operator.
This interaction Hamiltonian results  (up to some phase, which can be compensated easily) in the desired state, equivalent to applying the operator $\hat{U}_{int}(\theta)$ with $\theta=\hbar \frac{g^2}{\delta} \alpha^2 t_{int}$ where $t_{int}$ denotes the interaction time.
However, it is demanding to achieve a sufficiently strong nonlinear interaction corresponding to a $\theta$ of the order of $\frac{\pi}{2}$.
Therefore, here we shall also consider the case where $\theta$ is small (corresponding to a weak nonlinear interaction), similar to the analysis in Ref. \cite{hybridquantumrepeater}. 

A different approach was considered in the recent experiment of Ref. \cite{rempe_cat}, where a resonant light-atom interaction was employed in a cavity. 
More precisely, in this case the internal state of an atom determines
whether a light mode initially in a coherent state couples with the cavity or not.
In one atomic state (uncoupled with the cavity), the cavity mode and the incoming light pulse are on resonance
such that the light will enter the cavity and experience a $\pi$-phase shift
after leaving it again. In the other atomic state coupled with the cavity,
the effective cavity mode is no longer on resonance with the incoming pulse.
In this case, the light will not enter the cavity and immediately be reflected back
directly by the cavity mirror with no resulting phase shift.
As a consequence, an atomic superposition state leads to a state for the reflected pulse
that is entangled with the atom, similar to Eq. (\ref{eq:hybridrepstate1}), with a phase difference of $\pi$ for the two
coherent states.
 Therefore, in this case it is also possible to obtain $\theta=\frac{\pi}{2}$.

\section{\label{sec:quantumrepeater} Memory-Assisted Phase-Matching QKD protocol}

\subsection{Description of the protocol}
\begin{figure}
	\subfloat[\label{sfig:tf}]{
		\includegraphics[width=0.2\textwidth]{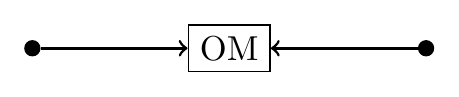}
	}
	\subfloat[\label{sfig:1segment}]{
		\includegraphics[width=0.2\textwidth]{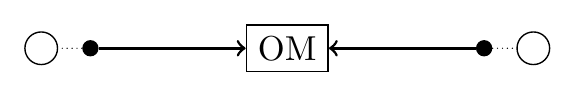}
	}\\
	\subfloat[\label{sfig:2segment}]{
		\includegraphics[width=0.4\textwidth]{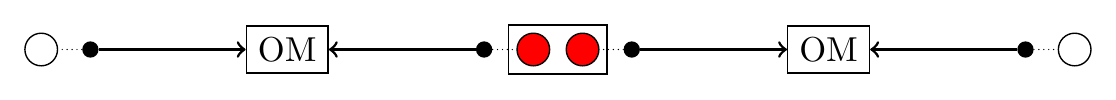}
	}\\
	\subfloat[\label{sfig:3segment}]{
		\includegraphics[width=0.4\textwidth]{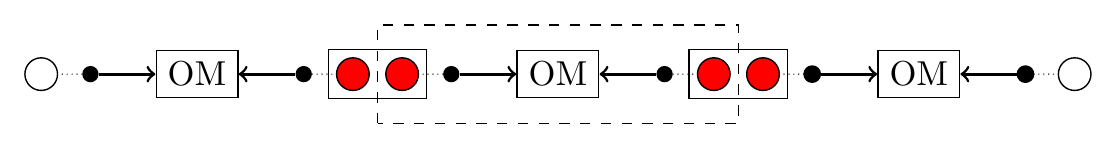}
	}\\
	\subfloat[\label{sfig:opticalmeasurement}]{
		\includegraphics[width=0.2\textwidth]{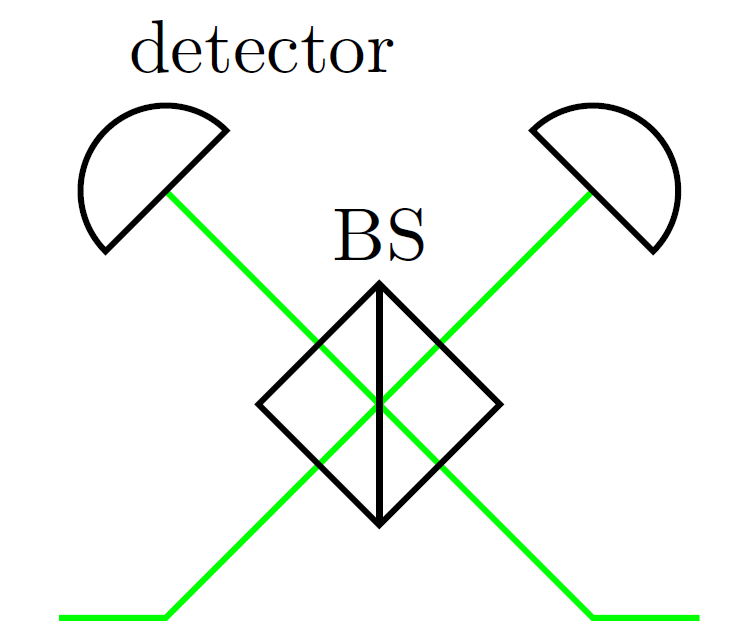}
	}
	
\caption{(Color online) Illustration of the protocol. (a) Phase-matching QKD. Alice and Bob send optical coherent states (black filled points) to Charlie who performs an optical measurement (OM).
(b) Entanglement-based variation of phase-matching  QKD ($n=1$). Alice and Bob each have an optical mode (black filled point) entangled with a short-lived memory (white filled circle). The optical fields are sent to Charlie's OM. The memories can be short-lived since it does not matter when Alice and Bob perform the measurements on their memories (as long as they wait with communicating their choice of measurement basis). (c) Two-segment HQR variant ($n=2$). Two copies of (b) are used where the memories in the central node need to be long-lived (red filled circles), since either of them has to wait until the other segment succeeds. When both segments succeeded, a Bell measurement is performed on the two long-lived memories for entanglement swapping. (d) Three-segment HQR variant ($n=3$). In order to obtain the $n$-segment repeater one simply needs to use $n-2$ inner segments (marked by the dashed box). Such a $n$-segment quantum repeater scheme consists of $2n$ physical segments.  (e) Set-up of the OM. Usually the detectors are on/off-detectors, but we could also use PNRDs. For $\theta\ll 1$ we only need one detector. `BS' stands for beam splitter. }
\label{fig:protocol}
\end{figure}

Let us start by describing the smallest example of our version of a HQR which is very similar to an entanglement-based description of phase-matching QKD (see Fig.\ref{fig:protocol} (a) and (b)).\\
 1) Alice and Bob each have an atom as a quantum memory and  generate a hybrid entangled state between their memory and an optical mode starting in a coherent state, resulting in $\frac{1}{\sqrt{2}}\left( \ket{\uparrow,\alpha \exp(-i\theta)}+\ket{\downarrow,\alpha\exp(i\theta)}\right)$.
Notice that Alice and Bob can also prepare BB84-states (thus distributing effective entanglement) instead of real entanglement. 
This is equivalent to the case where they generate real entanglement and perform measurements on the memories before sending the optical modes, because the measurements commute with Eve's operations provided that Alice and Bob only send information about the chosen measurement basis after establishing the raw key.
Whenever Alice or Bob should apply Pauli operations to their memories, but they already measured them, then this can be done via classical post-processing of the measurement data.
	 	The generation of these entangled states was described in the previous section.
		We will show that for our repeater protocol we can use, in principle, any $\theta>0$ at the expense of a larger amplitude $\alpha$ of the coherent state.
		Choosing a small $\theta$ is  also accompanied by the need of a better phase stabilization.\\ 
2) Alice and Bob send the optical part of their hybrid entangled states through a lossy channel of transmittance $\sqrt{\eta}$ to a middle station operated by Charlie ($\eta_\text{total}=\eta$).\\
3) If Charlie is honest, he applies a 50/50 BS to the two incoming optical modes with annihilation operators $\hat{a}$ and $\hat{b}$ described by the transformation, $\hat a' = (\hat a + \hat b)/\sqrt2$, $\hat b' = (\hat a - \hat b)/\sqrt2\,.
$
		Then he measures mode $b'$ with an on/off-detector or, alternatively, with a PNRD, while he does not need to measure anything for mode $a'$ (see Fig. \ref{fig:protocol} (e)).
		If he measures at least one photon, his measurement correlates Alice's and Bob's quantum memories.
\\
In order to distribute entanglement over very large distances, we divide the overall channel that connects Alice and Bob  into $n$ smaller segments where in each we run the above protocol.
The smallest example above then was for $n=1$ (Fig. \ref{fig:protocol} (b)) and the $n=2$-case with two repeater segments, each with a detection station in the middle (so, effectively four physical segments), can be seen in Fig. \ref{fig:protocol} (c).
As the next step, we perform entanglement swapping between neighboring quantum memories as soon as they are ready, as usual in quantum repeaters. In the end we have an (effective) two-qubit state shared by Alice and Bob that can be used for generating a secret key by employing e.g. the (entanglement-based) BB84 protocol.

Let us now get some intuition why we may use any $\theta>0$, especially $\theta\neq \frac{\pi}{2}$, and only need to measure one mode.
For this we will still omit channel losses.
We again consider the smallest $n=1$ case, corresponding to one repeater segment in the notation of general $n$.
The state before the BS is given by $\frac{1}{2}\left(\ket{\uparrow,\alpha \exp(-i\theta)}+\ket{\downarrow,\alpha\exp(i\theta)}\right)^{\otimes2}$. After the BS (and changing order) the state is given by \begin{align} \label{eq:state} \frac{1}{2}\left(\ket{\uparrow,\uparrow,\sqrt{2}\alpha e^{-i\theta},0}+\ket{\downarrow,\downarrow,\sqrt{2}\alpha e^{i\theta},0}\nonumber \right. \\  \left. +\ket{\uparrow,\downarrow,\sqrt{2}\alpha\cos(\theta),-i\sqrt{2}\alpha\sin(\theta)} \right. \\ \left. \nonumber+\ket{\downarrow,\uparrow,\sqrt{2}\alpha\cos(\theta),i\sqrt{2}\alpha\sin(\theta)}\right)\,,\end{align}
where the last two entries in each ket vector refer to the two modes $a'$ and $b'$, respectively.
In this simplified scenario, also assuming that Charlie uses a PNRD, by detecting mode $b'$ he projects the memories onto $\ket{\Psi^\pm}=\frac{1}{\sqrt{2}}(\ket{\uparrow,\downarrow}\pm\ket{\downarrow,\uparrow})$ where the sign depends on whether he measured an even or odd non-zero number of photons.
If we set $\theta=\frac{\pi}{2}$, we could in addition also use a PNRD for mode $a'$ and depending on the non-zero measurement outcome (even or odd number) Charlie's measurement would project the quantum memories onto $\ket{\Phi^\pm}=\frac{1}{\sqrt{2}}(\ket{\uparrow,\uparrow}\pm\ket{\downarrow,\downarrow})$.
 As a consequence, our wish to need only small $\theta$ comes at the price that the success probability is only half of the ideal probability of success for $\theta=\frac{\pi}{2}$.
 The protocol succeeds when there is at least one photon measured in mode $b'$ and therefore the success probability is given by $\frac{1}{2}\left(1-e^{-2\alpha^2\sin[2](\theta)}\right)$. 
When considering on/off detectors instead of PNRDs one projects onto a mixture of two Bell states.
Note that the post-measurement memory state and the success probability only depend on the product $\alpha\sin(\theta)$ and therefore we can use an arbitrarily small $\theta$ by employing correspondingly large amplitudes $\alpha$  in this simplified model.

\subsection{Channel-loss only}

As the next step we will include the lossy channel with transmittance $\sqrt{\eta}$ (between Alice/Bob and the middle station, again considering the $n=1$ case) and obtain the density operator of Alice's and Bob's qubits after Charlie's successful measurement.
 In order to keep this straightforward calculation clear, we will introduce auxiliary modes such that the lossy channel acts as a unitary operation on a larger Hilbert space. 
After Charlie's measurement we trace out all subsystems except Alice's and Bob's memory qubits. 
More details on this calculation can be found in App. \ref{app:calc}.
When  Charlie uses a PNRD the resulting density operator is given by 
\begin{align}
&\frac{1}{2}(1+ e^{-2(1-\sqrt{\eta})\alpha^2\sin[2](\theta)})\ket{\Psi^\pm}\bra{\Psi^\pm}\\ \nonumber+&\frac{1}{2}(1- e^{-2(1-\sqrt{\eta})\alpha^2 \sin[2](\theta)})\ket{\Psi^\mp}\bra{\Psi^\mp}\,,
\end{align}
where the upper sign holds in the even and the  lower sign holds in the odd photon number case. Due to the successful measurement the qubits can only be in the $\{\ket{\uparrow,\downarrow},\ket{\downarrow,\uparrow}\}$ subspace.
If Charlie uses an on/off detector, the density operator is given by
\begin{align}&\frac{1}{2}(1+ e^{-2(2-\sqrt{\eta})\alpha^2\sin[2](\theta)})\ket{\Psi^-}\bra{\Psi^-}\\ \nonumber+&\frac{1}{2}(1- e^{-2(2-\sqrt{\eta})\alpha^2\sin[2](\theta)})\ket{\Psi^+}\bra{\Psi^+}\,.
\end{align}
Here, the state $\ket{\Psi^-}$ has a larger probability because of the larger fraction of an odd non-zero photon number than that for an even non-zero photon number.
Therefore, Alice and Bob could exploit this to distill $1-h(\frac{1}{2}(1+e^{-2(1-\sqrt{\eta})\alpha^2\sin[2](\theta)}))$ or  $1-h(\frac{1}{2}(1+e^{-2(2-\sqrt{\eta})\alpha^2\sin[2](\theta)}))$ ebits in the cases of PNRDs or on/off detectors, respectively, using one-way classical communication in the asymptotic limit, where $h(\cdot)$ denotes the binary entropy function.
When using on/off detectors one obtains an ebit rate of
\begin{align}
\frac{1}{2}\left(1-e^{-2\sqrt{\eta} \alpha^2\sin[2](\theta)}\right)\left(1-h\left(\frac{1}{2}\left(1+e^{-2(2-\sqrt{\eta})\alpha^2\sin[2](\theta)}\right)\right)\right)\nonumber \\
\stackrel{\sqrt{\eta}\ll1}{\approx}\sqrt{\eta}\alpha^2\sin[2](\theta)\left(1-h\left(\frac{1}{2}\left(1+e^{-4\alpha^2\sin[2](\theta)}\right)\right)\right)\,.\nonumber\\
\end{align}
Note that this is the same as the secret-key rate of BB84 in the asymptotic limit.
The trade-off of the original HQR (assuming small $\theta$) between high fidelities for small $\alpha \theta$ and high success probabilities for large $\alpha \theta$ in the version with unambiguous state discrimination \cite{hybridquantumrepeater_usd} now becomes manifest in a high secret-key fraction (2nd factor) for small $\alpha \theta$ and a high raw rate (1st factor) for large $\alpha \theta$. However, the crucial difference is that the entanglement distribution probability in a single repeater segment ($n=1$) now scales with $\sqrt{\eta}$ instead of $\eta$ due to the middle station between Alice and Bob. 
 Since a similar expression appears in the PNRD case, it is useful to optimize the function  $f(x)=x\left(1-h(\frac{1}{2}(1+e^{-2x}))\right)$ and choose $\alpha^2\sin[2](\theta)$ accordingly.
 The maximum of $f$ is approximately $7.141\times10^{-2}$ with $x\approx0.229$. 
With this function, it can be seen easily that the use of PNRDs instead of on/off detectors only gives a factor of two improvement for the rate in the high-loss regime.
 Therefore, we will only consider on/off detectors since these are readily available in comparison to PNRDs.
The resulting overall ebit rate (allowing for small $\theta$) is given by $0.5 \times7.141\times10^{-2}\sqrt{\eta_\text{total}}$ (similar to \cite{twinfield_luetkenhaus}, \footnote{The difference in the protocol between \cite{twinfield_luetkenhaus} and our work with $n=1$ is that the authors of \cite{twinfield_luetkenhaus} use $\theta=\frac{\pi}{2}$ and two on/off detectors, such that their raw rate is larger by a factor of two. However, there are also differences in the approach of calculating the secret-key rate. We employ a BB84-like protocol since it easily allows us to go to a larger number of repeater segments, whereas \cite{twinfield_luetkenhaus} consider the `Devetak-Winter formula' for obtaining the secret-key fraction by calculating the mutual information between Alice's and Bob's bits and estimating the mutual information between Eve and the key via the Holevo information. This approach allows the authors of \cite{twinfield_luetkenhaus} to employ only coherent states for estimating Eve's information, while in our approach we need to generate hybrid entangled or cat states, even for the simplest $n=1$-case without memory assistance.}).

Next we consider the case of $n$ segments (see Fig. \ref{fig:protocol} (d)).
 It is then straightforward to calculate Alice's and Bob's density operator after the quantum teleportation steps, because the input states are Bell-diagonal (see App. \ref{app:pauli} for details).
For the case of on/off-detections, up to suitable Pauli operations (which can also be applied via classical post-processing if Alice and Bob already measured their qubits in the beginning) after the Bell measurements on the memory qubits for entanglement swapping (see Fig. \ref{fig:protocol} (c) for the $n=2$ case), Alice and Bob share the (effective) state 
\begin{align}
&\frac{1}{2}(1+ e^{-2n(2-\sqrt{\eta})\alpha^2\sin[2](\theta)})\ket{\Phi^+}\bra{\Phi^+}\\ \nonumber+&\frac{1}{2}(1- e^{-2n(2-\sqrt{\eta})\alpha^2\sin[2](\theta)})\ket{\Phi^-}\bra{\Phi^-}\,.
\end{align}  When using PNRDs one obtains a similar state with a different coefficient of $\ket{\Phi^\pm}$ ($1-\sqrt{\eta}$ instead of $2-\sqrt{\eta}$).
Let us consider a scheme where we try to distribute the entanglement in all segments in parallel and only at the end we perform the entanglement swapping everywhere. 
Using the results for the exact raw rate with deterministic entanglement swapping \cite{hybridrepeaterrateanalysis}, one can calculate the obtainable ebit/secret-key rate for this simple case exactly.
 However, to obtain a rough overview it is useful to apply an approximation for the raw rate (assuming $\sqrt{\eta}\ll1$, see details in App. \ref{app:approx}) and use the optimal value for $n \alpha^2 \sin^2(\theta)$, resulting in an overall secret-key rate of 
 \begin{equation}
 \sqrt[2n]{\eta_\text{total}}H(n)^{-1}\frac{0.07}{2n} \sim 3.57\times10^{-2} \frac{\sqrt[2n]{\eta_\text{total}}}{n\left(\gamma+\ln(n)\right)}\,,
 \end{equation}
  where $H(n)$ are the harmonic numbers and $\gamma=0.57721\dots$ is the Euler-Mascheroni constant.
Notice that we always have to reduce the mean photon number $\alpha^2$ of each optical pulse with increasing $n$ $(\alpha_\text{optimum}\approx\frac{1}{\sin(\theta)}\sqrt{\frac{0.229}{2n}})$.
All these considerations are for secret-key rates per channel use (and per mode, but in our schemes, the optical states are single-mode). We define one channel use as a single attempt to \text{generate entanglement in all} repeater segments.

One benefit of this scheme is that in order to obtain a secret-key rate scaling of $\sqrt[2n]{\eta_\text{total}}$ one only needs $n-1$ stations equipped with quantum memories.
In comparison, a standard quantum repeater \cite{repeater_briegel}\footnote{Note that there are well-known proposals for quantum repeaters that are based on single-photon interference and thus intrinsically contain the twin-field-type scaling advantage.
One such protocol makes use of a single atom or spin entangled
with a light mode that either contains a photon or not \cite{cabrillo_repeater}, see also \cite{wehner2}.
Another protocol, proposed by Duan, Lukin, Cirac, and Zoller (DLCZ) \cite{DLCZ}, initially employs entanglement between
a light mode and the collective spin mode of an atomic ensemble.
The finally resulting two-mode single-excitation spin entanglement in DLCZ,
however, cannot be straightforwardly used for applications like QKD
and therefore DLCZ suggest a postselection strategy by considering
two copies of a repeater chain and accepting only those cases
where each end point of the double-chain state carries exactly one spin excitation.
As a consequence, the DLCZ scheme loses its additional scaling advantage.
The schemes of Refs.\cite{wehner2,cabrillo_repeater} do not share this complication,
because their resulting spin-spin entanglement is of immediate use.} would need $2n-1$ stations with memories when directly employed for QKD with Alice and Bob  immediately measuring their qubits (otherwise the standard repeater uses $2n+1$ memories, while our scheme would use $n+1$ memories). 
Note that the scaling of $\sqrt[2n]{\eta_\text{total}}$ is consistent with the ultimate end-to-end capacity in repeater-assisted quantum communication where the channel is divided into $2n$ physical channel segments (assuming large segment lengths) \cite{pirandola_repeater}.
When considering first experimental realizations of small-scale memory-based quantum repeaters, using a scheme like ours (or related schemes like those of Ref. \cite{wehner2}) could be beneficial, because in order to obtain a secret-key rate scaling of $\sqrt[4]{\eta_\text{total}}$ only a single memory station is needed instead of three. 

For the case of this section where channel loss is the only error considered, the distillable entanglement (when allowing one-way, forward classical communication)  coincides with the asymptotic secret-key rate obtainable with BB84.
In order to obtain a reasonably realistic description of such a repeater, we also have to include dark counts and the efficiency of the on/off detectors, memory dephasing, phase mismatch, and errors in the deterministic entanglement swapping which will be described by a depolarizing channel.
Before turning to such a model including all of these errors, however, one may first only include the most important errors which still enables one to see their influence onto the secret-key rate in simple, analytical expressions.
For our treatment here, all conceptual and technical details regarding the more realistic cases beyond just channel loss are presented in the appendices. There we first consider detector inefficiencies and memory dephasing where we can still describe the resulting states as mixtures of two Bell states.
Including detector efficiencies ($\eta_{det}$) is trivial, because we only have to substitute $\sqrt{\eta}\rightarrow \sqrt{\eta}\times \eta_{det}$.
However, things become trickier when considering the dephasing in the memories. 
Nonetheless, since the dephasing channel is a Pauli channel, it commutes with the entanglement swapping and therefore we can assume that we first distribute perfect entanglement via multiple quantum teleportations and then apply the errors to the qubits (according to the loss channel and the memory dephasing, see App. \ref{app:model},\ref{app:pauli}).
Later we also consider imperfections of the Bell measurements which still result in Bell-diagonal states.
Finally, we will also take into account dark counts, eventually leading to Bell-non-diagonal states. A detailed discussion of the influence of these error sources to the secret-key rate is given in App. \ref{app:discuss}.
We also present a detailed discussion on the use of homodyne detectors for our scheme in App. \ref{app:discuss},\ref{app:homodyne}. The secret-key rates as obtainable with our model (based on on/off detectors) will be presented and compared among different scenarios in the following section. The extra experimental parameters as required for the discussion there are all introduced in App. \ref{app:model},\ref{app:discuss}.

\section{\label{sec:comparison} Comparison of secret-key rates}
\subsection{Secret-key rate per channel use}
Let us now consider the performance in terms of BB84 secret-key rates per channel use of our proposed scheme for some physically reasonable parameters.
We start with the example of a two-segment repeater (i.e., $n=2$, corresponding to two segments connected at a memory station and each segment equipped with an optical middle station, see Fig. \ref{fig:protocol}(c)).
We assume the following parameters (similar to Ref. \cite{twinfield_luetkenhaus}): 
\begin{itemize}
\item $\sqrt{\eta}=0.15\exp(-\frac{L}{2n L_{att}})$
\item $L_{att}=22\text{km}$
\item $\alpha=23.9$
\item $\theta=0.01$
\item dark count probability $p_{dark}=8\times 10^{-8}$
\item $p_{depol}=10^{-2}$
\item $\tau=\frac{L}{nc}$
\item $c=2\times10^8\frac{m}{s}$
\item error correction inefficiency $f_{EC}= 1.15$
\end{itemize}
The transmission parameter $\sqrt{\eta}$, when we set $n=2$, corresponds to a quarter of the total distance $L$ between Alice and Bob, because every mode travels only for this distance to the corresponding detector station.  This parameter also includes a finite detector efficiency (factor $p_{det}=0.15$).
We shall also consider perfect detectors, $p_{det}=1$.
Since we do not know the optimal value of $\alpha$ (for given $\theta$) when considering all possible errors in our model, we simply use the optimal $\alpha$ from the loss-only case assuming $n=2$. This already gives a good starting point for $\alpha$ which we use everywhere unless stated otherwise.
 Further parameters are explained in Appendices \ref{app:model} and \ref{app:discuss}.

The BB84 secret-key fraction \cite{plobhistory} is given by 
\begin{equation}
1-h(e_X)-f_{EC} h(e_Z)\,,
\end{equation}
where $e_{X/Z}$ are the error probabilities in the $X$- and $Z$- basis which can also be expressed in terms of the four Bell-state coefficients of the density matrix.
Note that we consider the biased BB84 scheme where one of the two bases is employed more often allowing to increase the sifting factor to unity in the asymptotic limit of infinite repetitions \cite{biasedBB84}.
The overall secret-key rate is then given by the product of the raw rate and the secret-key fraction.

\begin{figure}
\includegraphics[width=0.4\textwidth]{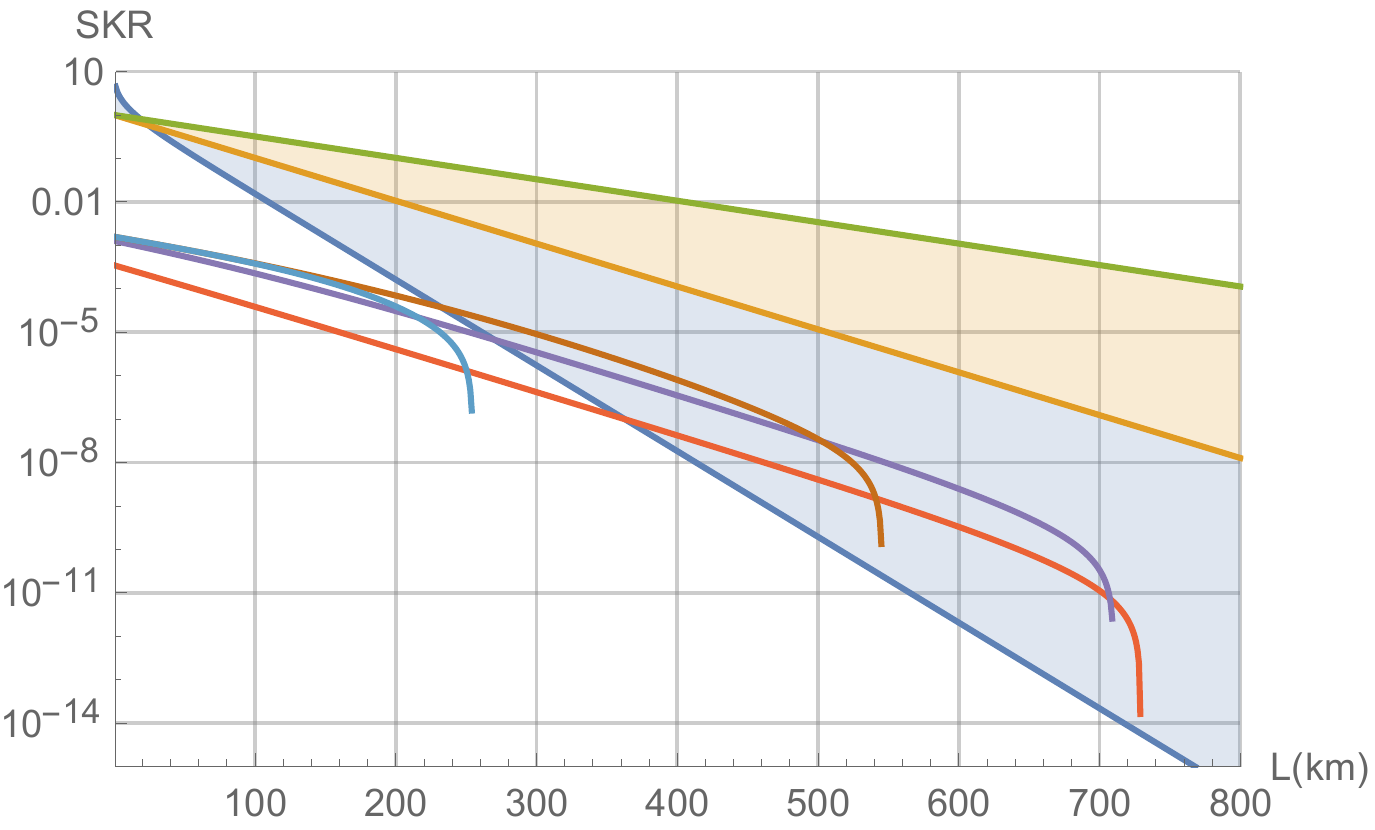}
\caption{(Color online) Secret-key rates for a two-segment repeater ($n=2$, parallel scheme) without phase mismatch assuming the parameters as listed in the main text (including $p_{\text{det}}=0.15$) and a memory coherence time $T$ of 1 second.  The straight lines (from bottom to top) denote the PLOB bound,$\sqrt{\eta_{total}}$, and $\sqrt[4]{\eta_{total}}$.  The rates are for different values of the memory cut-off (10,100,1000,10000) (from right to left). The areas between PLOB and $\sqrt{\eta_{total}}$ and between $\sqrt{\eta_{total}}$ and $\sqrt[4]{\eta_{total}}$ are highlighted in color.}
\label{fig:comparison_cutoff_1sec}
\end{figure}

\begin{figure}
\includegraphics[width=0.4\textwidth]{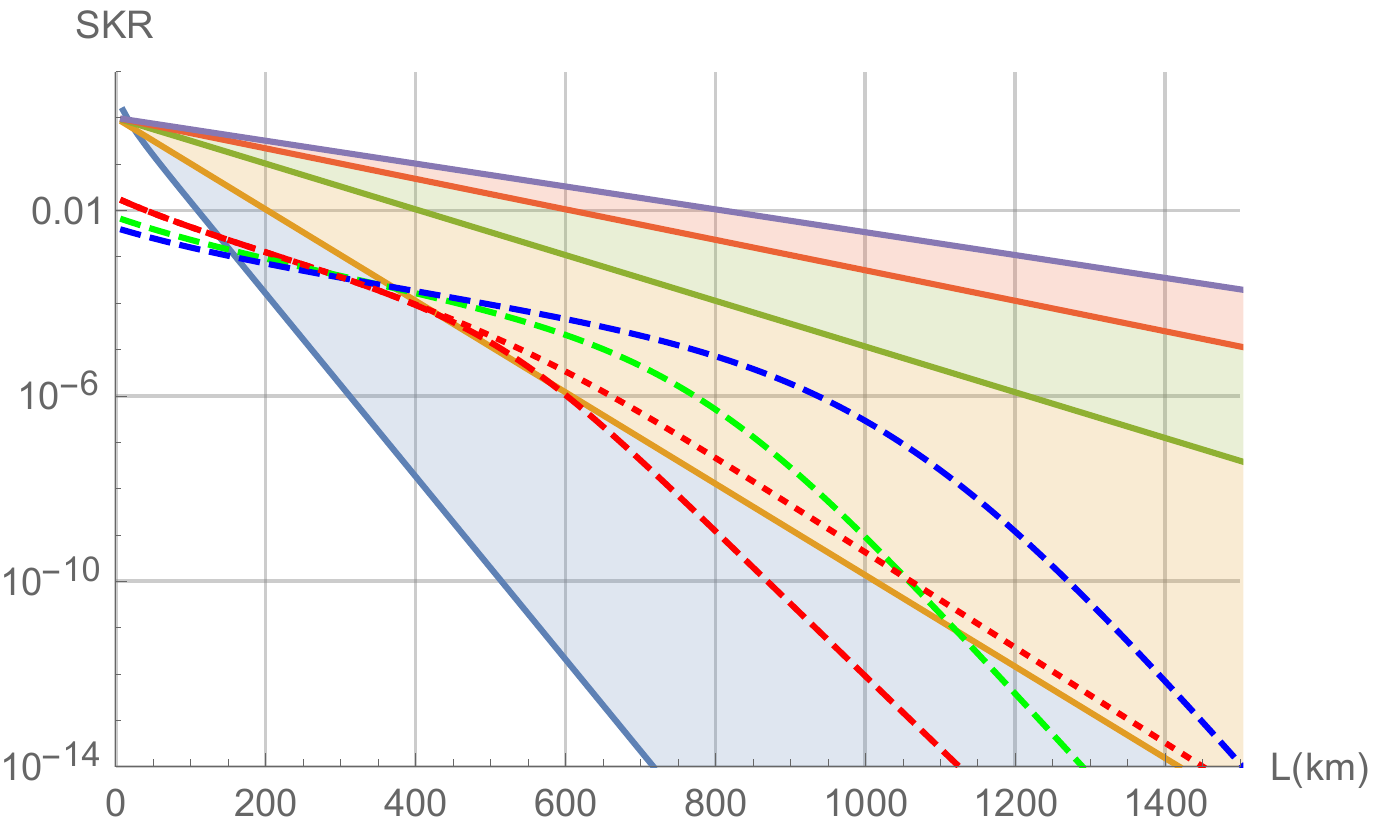}
\caption{(Color online) Secret-key rates for a repeater with $n$=2 (red), 3 (green), 4 (blue) (dashed, from left to right)  segments using a sequential protocol (parallel for $n=2$) without cut-off (dashed lines, $\alpha=23.9$ in all cases). For all curves we consider a finite memory coherence time of 10 seconds (no other errors are assumed). The red dotted line denotes a $n=2$ scheme where we do use a cut-off. The benchmarks (from bottom to top) PLOB, $\sqrt{\eta_\text{total}}$,  $\sqrt[4]{\eta_\text{total}}$,  $\sqrt[6]{\eta_\text{total}}$, and  $\sqrt[8]{\eta_\text{total}}$ can also be seen. The regions between two of those benchmarks are highlighted in color accordingly.}
\label{fig:crossover}
\end{figure}

The memory coherence time $T$ and the phase mismatch will be varied in order to assess their influence on the secret-key rate (see App. \ref{app:discuss}).
Let us first study the effect of the memory dephasing, since insufficient coherence times are an important issue for quantum repeaters.
As it can be seen in Figs. \ref{fig:comparison_memory} and \ref{fig:comparison_memory2} (in App. \ref{app:discuss}), one really needs demanding memory coherence times such as 1000 seconds or more in order to be able to expect nearly the total benefit of the memory-based repeater capabilities.
 When considering more realistic, currently available memories with a coherence time of around 1s \footnote{Currently available memory coherence times are ranging from several $\mu s$ (quantum dots) to tens of $ms$ (color centers, atoms, and ions) \cite{white_paper}. Although there are very recent reports on approaching coherence times of up to a few or even above 60 minutes \cite{Wang_memory}, we assume that future coherence times that are also compatible with the requirement of telecom-frequency conversion are within the range of almost 1ms (quantum dots) and 0.1-1s (atoms and ions) up to 10s (color centers). In our quantitative rate analysis including memory dephasing we will thus have a particular focus on coherence time values of 0.1s, 1s, and 10s	(see especially Figs. \ref{fig:comparison_cutoff_1sec}-\ref{fig:hybrid_comparison}). For a more detailed discussion on the interplay between experimental clock times (with or without the need of additional spin sequences on the memory qubits), memory coherence times, and the need for frequency conversion, for various experimental hardware platforms, see Ref. \cite{white_paper}. In that reference as well as in the present work, the focus is on single-spin quantum memories subject to dephasing rather than spin ensembles (as employed in atomic-ensemble-based quantum repeaters \cite{DLCZ}), which instead must be modelled including memory loss acting on collective, bosonic spin modes.}, it can be seen that it is not even possible to overcome the PLOB bound (Fig. \ref{fig:comparison_memory} with inefficient detectors).
 This means in this case the additional memory element even worsens the secret-key rate in comparison to simple twin-field QKD.
However, we also found that the detection efficiency $p_{det}$ is a highly influential parameter determining whether PLOB can be exceeded or even the ultimate $\sqrt[4]{\eta_{total}}$-scaling can be approached, with realistic ($\sim 1s$) or potential future ($\gtrsim10s$) coherence times, respectively (see Fig. \ref{fig:comparison_memory2})\footnote{Throughout all plots, as benchmarks, we show the PLOB bound $-\log_2(1-\eta_{total})$ for point-to-point communication between Alice and Bob \cite{PLOB} and, instead of the ultimate repeater bounds for a quantum repeater with $2n$ physical segments $-\log_2(1-\sqrt[2n]{\eta_{total}})$ \cite{pirandola_repeater}, several benchmarks of the form $\sqrt[2n]{\eta_{total}}$ (which up to a factor of 1.44 coincide with the former for small $\sqrt[2n]{\eta_{total}}$), since our particular qubit-based scheme can never exceed $\sqrt[2n]{\eta_{total}}$ similar to the ideal standard twin-field scheme that never goes above $\sqrt{\eta_{total}}$.}.

Based on the above observations one may infer that the MA-PM QKD scheme cannot help increasing long-distance secret-key rates using currently available memories and finite, modest detector efficiencies.
However, up to now we assumed that the participants will always wait until the entanglement is distributed in both segments no matter how long this distribution lasts for.
It is  possible though to introduce a maximal memory waiting time \cite{wehner,wehner2,jiang_cutoff,rawratecutoffCollinsPRL,onthewaitingtime,Praxmeyer_finite_memory} until which the entanglement must be distributed in both segments, otherwise the entanglement already distributed in one segment is discarded in order to prevent large error rates at the expense of a lower raw rate.
References \cite{rawratecutoffCollinsPRL,onthewaitingtime} derive the raw qubit rate for a two-segment repeater with such a memory cut-off, while Ref. \cite{Praxmeyer_finite_memory} presents a rate formula for the more general case of arbitrarily many segments under the constraint of deterministic entanglement swappings.
 References \cite{wehner,wehner2} analyze the dephased qubit states for schemes with at most two segments.
 As it can be seen in Fig. \ref{fig:comparison_cutoff_1sec} it is possible to overcome the PLOB bound by introducing a cut-off and, furthermore, it is even possible to distribute secret keys over a distance of 700 km and more with modestly performing memories and detectors (compare this with Fig. \ref{fig:comparison_memory}, even with $T=\infty$).
In this work we only consider rates including cut-off for $n=2$. 

We expect that a cutoff will also enhance the final rates for more than two segments. Thus, our rate analysis leads us to the following conclusion. Even though the PLOB bound can in principle be exceeded for our $n=2$ scheme by introducing a memory cutoff, a higher experimental cost would be needed, i.e. sufficiently efficient memories and detectors, in order to benefit from the improved scaling of our $n=2$ scheme compared with twin-field QKD. However, in Sec.IV.B, we will see that when rates per second are considered, it is generally hard for a small-scale repeater like our $n=2$ scheme to compete against twin-field QKD at high clock rates. Therefore, we will also consider more than two repeater segments (as for an alternative, we also explore the possibility of an asymmetric two-segment repeater operating at a higher clock rate in App. \ref{sec:asym}).

In Fig.~\ref{fig:crossover} one can see the scaling behavior of repeaters based on our protocol with $n=2,3$ or even 4 repeater segments considering a finite memory coherence time of 10 s and no additional errors in comparison to the PLOB bound and ideal quantum repeaters. For all $n$ we choose $\alpha=23.9$ even though it is generally not the optimal value in the loss-only case, but it yields better rates when considering other errors. However, note that we did not try to find an optimal $\alpha$ in the general case. When we optimize $\alpha$ this will be explicitly stated. We found that for these three different segment numbers PLOB is overcome at an overall distance of approximately $140$ km. However, since the PLOB bound can be overcome by twin-field QKD without memory stations, the more relevant benchmark for our protocol may be $\sqrt{\eta_\text{total}}$ which can be exceeded at approximately 350 km. Due to the coherence time of only 10 s we can barely surpass $\sqrt{\eta_\text{total}}$, but with an appropriately chosen cut-off parameter (in the $n=2$ case) we can overcome this benchmark even for distances between 450 and 1500 km (see Fig. \ref{fig:crossover}). Furthermore, we find that by making use of a memory cut-off and perfect-efficiency detectors, but also including dark counts and an imperfect Bell measurement, it suffices to require a coherence time of 5s for overcoming $\sqrt{\eta_{total}}$ (not shown in plots). In order to obtain better rates than in the ideal twin-field scheme a coherence time of 1s suffices, even without making use of a memory cut-off (see Fig. \ref{fig:comparison_memory2}).

\subsection{Secret-key rate per second}

For practical applications, the secret-key rate per second is the more important figure of merit for comparing quantum repeaters with other types of QKD schemes. A large disadvantage of scalable memory-based quantum repeaters is that they rely on classical communication for confirming success, setting an upper bound on the repetition rate due to classical communication times. For example, when assuming a spacing of 100km between two repeater stations this limits the repetition rate to the order of kHz. However, theoretically, this also makes it easy to convert the secret-key rate per channel use to a secret-key rate per second, because the (classical and quantum) communication times are typically much larger than the local operation times and thus the latter can be neglected in the regimes that we mainly consider here \footnote{The problem of these low repetition rates can be circumvented when using so-called third-generation quantum repeaters which make use of quantum error correction \cite{generations}. However, an optical implementation of suitable quantum error correcting codes is currently still hard to achieve.}. Therefore, in order to perform better than twin-field QKD in terms of secret-key rate per second by using a memory-assisted repeater one needs to employ sufficiently many repeater stations for a given total distance (with $\eta_{total}\ll1$), such that the communication times become smaller (and also the scaling advantage increases). However, even for repeater spacings as small as 100m, the repetition rate only grows to the order of MHz. Hence, one can see that a scalable memory-based quantum repeater with a reasonable repeater spacing has to outperform twin-field QKD by many orders of magnitude in terms of secret-key rate per channel use, only to obtain rates similar to twin-field QKD per second.
 Nonetheless, there are at least three reasons for why it is still beneficial to employ our memory-assisted schemes.
 
  First, like general memory-based quantum repeaters, in principle, long-distance regimes become accessible for rates per second that are otherwise (including for twin-field QKD) unreachable at the same rates. This happens because of the scaling advantage which eventually dominates over the clock-rate disadvantage for sufficiently long distances. At such distances, the final rates per second are generally low, but this applies to both twin-field QKD and MA-PM QKD while rates end up strongly biased towards MA-PM QKD with growing distance. In this case, the small final rates of MA-PM QKD may be enhanced up to practical values by employing many repeater chains in parallel (multiplexing).
Second, also for distances where dark counts greatly reduce the secret-key rate a repeater can overcome the twin-field QKD secret-key rate per second. However, it is also possible to obtain the same effect by using entangled light sources with a high repetition rate as a relay in order to keep the dark count effect small. With our system, such a relay could be realized when all spins of the hybrid spin-light entangled states are measured immediately.  In this case, we only lose a factor of $\frac{1}{2}$ when employing small $\theta$, however, with a simple relay ($n=2$), as we no longer make use of memories, the effect is squared.
Third, unlike direct-transmission or twin-field QKD at high repetition rates, our memory-based schemes can also be used in applications different from QKD such as distributed quantum computing.

\begin{figure}
\includegraphics[width=0.4\textwidth]{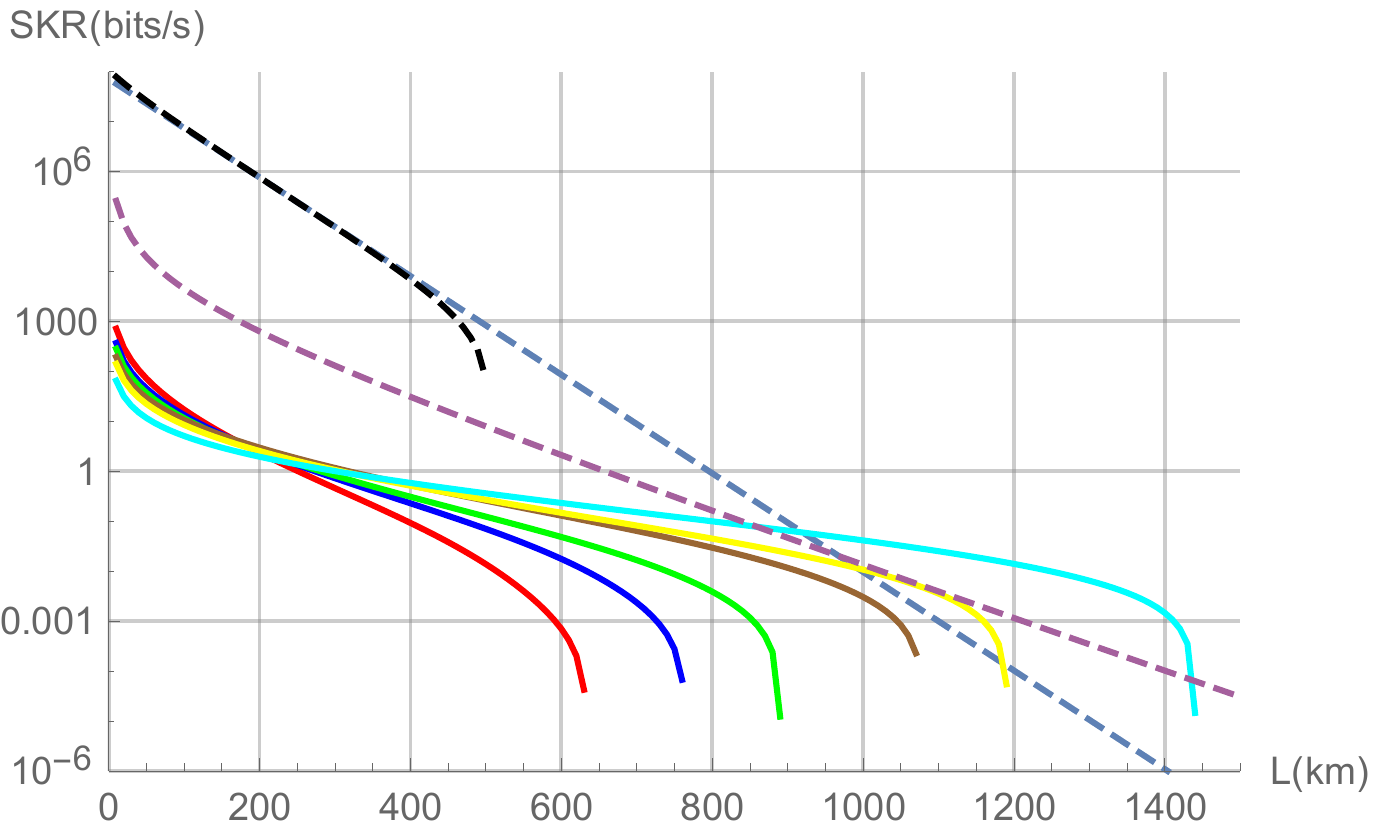}
\caption{(Color online)  Comparison of the secret-key rate in bits/s between twin-field QKD (ideal: blue dashed, with dark counts: black dashed) assuming a repetition rate of 1 GHz and our proposed scheme for $n=2,3,4,6,8,16$ (left to right in terms of vanishing rates, parallel scheme for $n=2$) where we assumed detectors with unit efficiency ($p_{det}=1$),  a dark count rate of $7\times 10^{-8}$, a memory coherence time of $10s$ and $p_{depol}=10^{-3}$. The dashed purple line (for $L\approx0$ beginning at a rate of $\sim10^5 \frac{\text{bits}}{\text{s}}$) represents an ideal standard repeater with four physical segments attaining the repeater-assisted capacity \cite{pirandola_repeater} whose repetition rate is limited by the communication time. Notice that for $n$=8 and distances as large as 1000km we outperform ideal twin-field QKD with a noisy repeater in terms of secret-key rate per second while still attaining rates as high as $10^{-2}$ Hz without making use of memory cut-offs.} 
\label{fig:ideal_skr_per_sec}
\end{figure}

As it can be seen in Fig. \ref{fig:ideal_skr_per_sec} our proposed schemes (for $n\geq8$) can outperform (in terms of secret-key rate per second) idealized twin-field QKD even when we consider dark counts, memory dephasing ($T=10s$) and depolarizing errors $p_{depol}=10^{-3}$ in our repeater scheme for distances above 1000km \footnote{In Fig. \ref{fig:ideal_skr_per_sec} we need to optimize $\alpha$ for the different schemes. Otherwise, if we use $\alpha=23.9$ in all schemes the rates for $n=16$ become the worst in the plots. We used the following values for $\alpha$:  30,23.9,23.9,18,17,9 (ordered by increasing $n$).}. However, even a maximally idealized four-segment quantum repeater (in a standard approach employing as many memories as our $n=4$ scheme) that attains the corresponding repeater capacity  \cite{pirandola_repeater}  outperforms idealized twin-field QKD just at distances of 1000km. Thus, it is a rather fundamental problem to overcome idealized twin-field QKD with further scalable memory-based quantum repeaters for a small number of memories in a regime where the single-chain secret-key rate per second is not too low for practical purposes. Also notice that here we did not consider a memory cut-off (for $n>2$ there are many different strategies to implement such a cut-off protocol) and therefore we expect that it is possible to improve the secret-key rates of our schemes significantly (recall the improvement in the comparison between Fig. \ref{fig:comparison_cutoff_1sec} and Fig. \ref{fig:comparison_memory}). When comparing our schemes with a noisy twin-field QKD protocol it is easy to see that our schemes allow for a longer communication range until the secret-key rates drop to zero.

\begin{figure}
\includegraphics[width=0.4\textwidth]{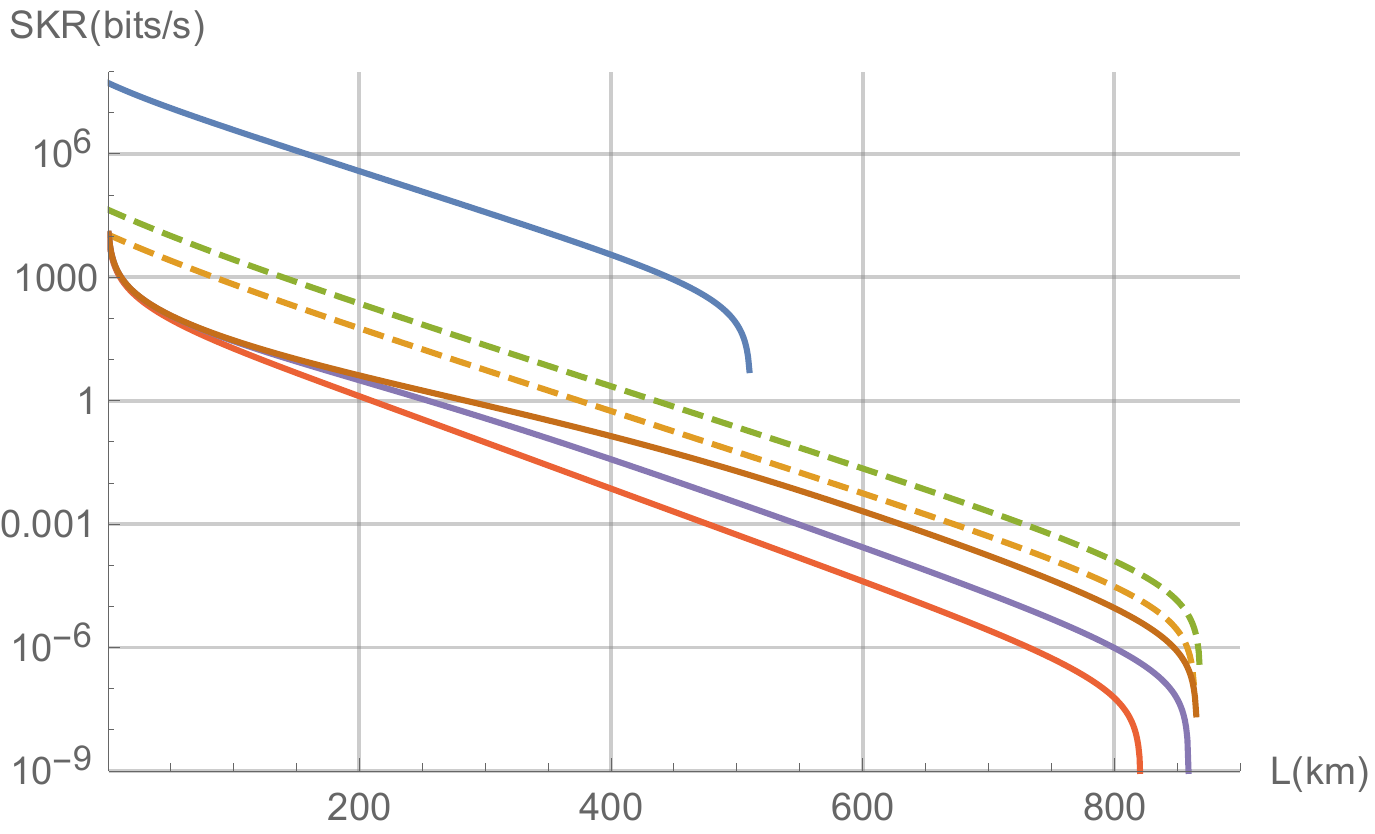}
\caption{(Color online)  Comparison of the secret-key rate in bits/s between twin-field QKD (blue, top) assuming a repetition rate of 1 GHz and our proposed scheme for $n=2$   including a memory cut-off and assuming different memory coherence times of (0.1,1,10) seconds (solid lines, bottom to top) ($p_{det}=1$, parallel scheme, other parameters are as described in the main text). The dashed lines (yellow: small $\theta$\, hence smaller rate) refer to a relay configuration assuming a repetition rate of 1 MHz taking into account the finite spin-light interaction times for the optical entangled-state generations in our relay.}
\label{fig:secret_key_per_sec_cuttoff}
\end{figure}

In Fig. \ref{fig:secret_key_per_sec_cuttoff} it can be seen that our scheme with $n=2$ including memory cut-off is able to outperform twin-field QKD in a scenario where dark counts are taken into account. Even a scheme with memories of rather low coherence time such as 0.1s is able to outperform realistic twin-field QKD at a distance of approximately 440km, though resulting in a rather low secret-key rate of 10 bits/h. Memories with such coherence times are already available \cite{white_paper}. However, it can also be seen that a similar enhancement is achieved with a relay (which actually scales better than the memory-based scheme for a coherence time of 0.1s). In order to see an improved scaling for the repeater, one needs a coherence time as large as 10s. Since the huge gap between twin-field QKD and our proposal in terms of secret-key rate per second in some regimes originates from the different repetition rates of both schemes, it is reasonable to consider the possibility for $n=2$ not to place the beam splitter in the middle for one segment, but in an asymmetric way (thus modifying Fig. \ref{fig:protocol}(c)). Improvement is possible then, because Alice and Bob can send light states at an in principle arbitrarily high repetition rate, since they only need the information regarding success from the beam splitter in order to decide whether they should count or discard that round in the final classical post-processing. However, the memory station requires this information immediately in order to decide if the state in the memory should be held or discarded. When placing the beam splitter nearer to the memory, one decreases the secret-key rate per channel use, but at the same time enhances the possible repetition rate. We discuss this scheme in App. \ref{sec:asym}. We find that for a not fully asymmetric scheme one can increase the secret-key rate per second by up to a few percent. 

\subsection{Comparison with USD hybrid repeater}

Let us now compare our new HQR with a HQR that uses on/off detectors for unambiguous state discrimination (USD) \cite{hybridquantumrepeater_usd}.
In our scheme in each segment we have two qubit memories each interacting nonlinearly with a coherent state and these optical states are then sent to a middle station with a 50:50 beam splitter followed by an on/off detector. In the USD scheme we have two memories, but only one optical state. First this state interacts with the first memory, is then sent to the other, and interacts with this second memory. In the end, a USD measurement using linear optics, phase-space displacements, and three on/off detectors is performed. Thus one can see that both schemes employ very similar resources. We can evaluate and compare the two schemes in a simple error model where we consider channel loss, depolarization and memory dephasing.

\begin{figure}
	\includegraphics[width=0.4\textwidth]{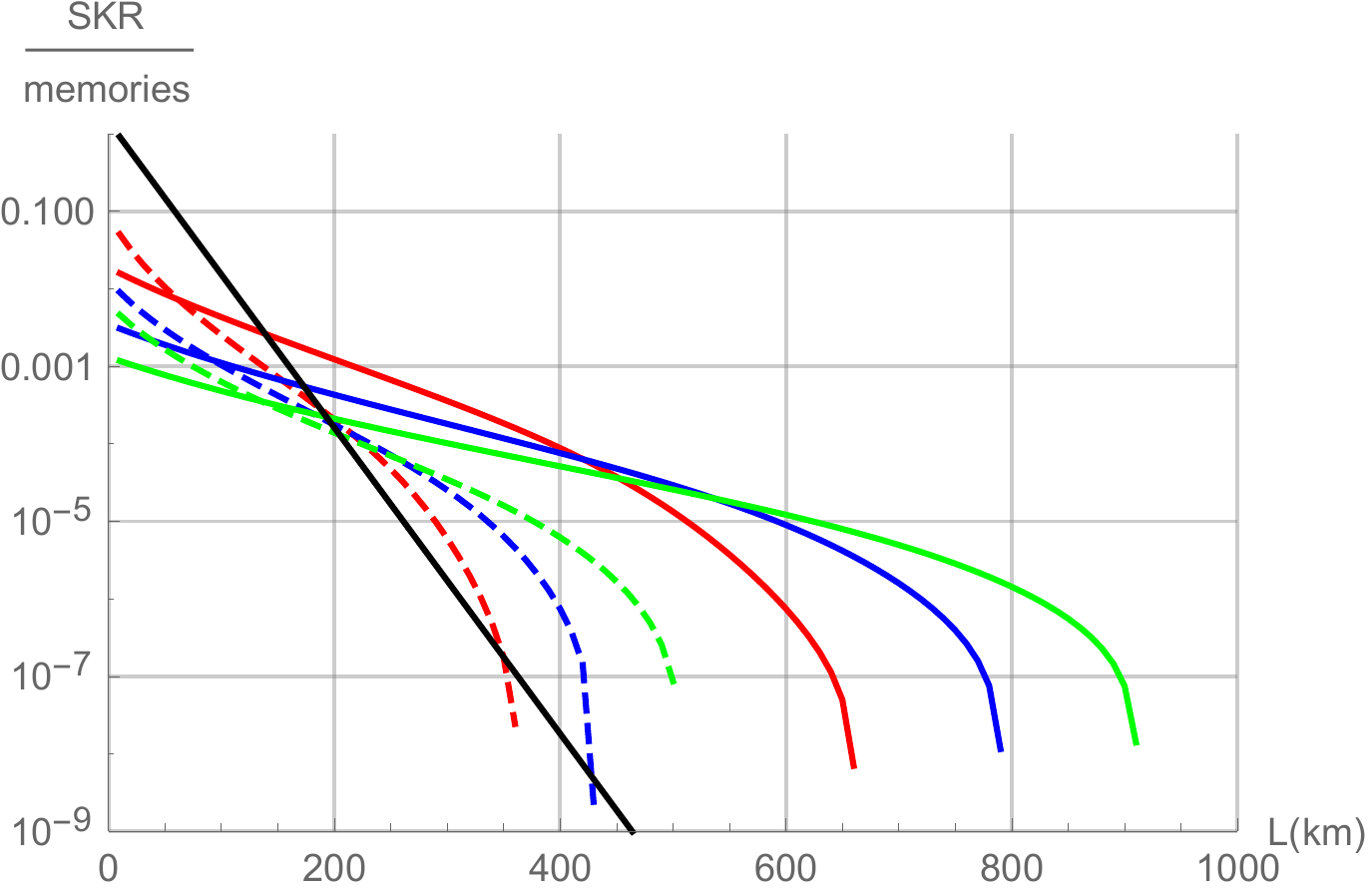}
	\caption{(Color online)  Comparison of the secret-key rate per channel use per employed memory (station) for our scheme (solid lines) and the USD hybrid scheme (dashed lines) for $n=2,3,4$ (from left to right in the regime of rates dropping towards zero) assuming a coherence time of $T=10$  $s$, a depolarizing channel with $p_{depol}=10^{-3}$ and a sequential scheme (parallel for $n=2$). The black solid line corresponds to the PLOB bound.}
	\label{fig:hybrid_comparison}
\end{figure}

In our scheme, the probability to generate entanglement in one segment in a single try is given by $\frac{1}{2}\left(1-e^{-2\sqrt{\eta} \alpha^2\sin[2](\theta)}\right)$, while for the USD hybrid repeater it is given by $\frac{1}{2}\left(1-e^{-2\eta \alpha^2\sin[2](\theta)}\right)$. Here we can already see the improvement of our scheme in the raw rate since $\eta$ is simply replaced by $\sqrt{\eta}$. The loss channel and the measurement also induce a dephasing channel with parameter $e^{-2(2-\sqrt{\eta})\alpha^2\sin[2](\theta)}$ in our scheme. In the USD scheme this is given by $e^{-2(1-\eta)\alpha^2\sin[2](\theta)}$. The memory dephasing works similar in both cases, but in our scheme the duration of a single entanglement generation attempt is given by $\frac{L}{nc}$, in the USD scheme by  $2\frac{L}{nc}$ \footnote{For the USD scheme it is also possible to shorten the duration of one time step to $\frac{L}{nc}$ by switching the role of sender/receiver. If the entanglement generation fails Bob usually communicates this failure to Alice who then tries again. However, briefly after sending the classical communication he can also start to send an optical pulse to Alice, who needs this short break for switching from sender to receiver mode which might be experimentally complicated.}. As it can be seen in Fig. \ref{fig:hybrid_comparison}, already our $n=2$ scheme  gives better secret-key rates per employed memory than the USD hybrid repeater for $n=2,3,4$ for relevant distances. In the regime of small distances the USD scheme achieves rates slightly better than our scheme, because for $\eta=1$ there is no dephasing due to loss and the measurement in the USD scheme. However, our scheme always has dephasing originating from the usage of the on/off detectors. Nonetheless, our scheme has a better distance scaling and therefore our schemes achieve better rates than the USD scheme for the relevant large-distance regime. For these distances our schemes often achieve rates which are orders of magnitude better then those of the USD scheme. Thus, we obtain a better secret-key rate while employing a smaller supply of quantum memories.

\section{\label{sec:conclusion} Conclusion}

We introduced a measurement-device-independent QKD scheme based on the twin-field QKD concept but making use of memories in order to extend the overall distance for which a secret key can be distributed. The secret-key rates per channel use of our scheme scale as  $(n H(n))^{-1}\sqrt[2n]{\eta_{total}}$ (harmonic number $H(n)=\gamma+\ln(n)+\mathcal{O}(n^{-1})$) in the loss-only case (assuming $\sqrt[2n]{\eta_{total}}\ll1$ and using a parallel entanglement distribution scheme), where $\gamma=0.57721\dots$ is the Euler-Mascheroni constant and $n$ is the number of repeater segments, each equipped with memory stations at their ends and a beam splitter and optical-detector station in their middles. 
The transmission parameter $\eta_{total} = \exp(-\frac{L}{L_{att}})$ represents the total channel connecting Alice and Bob separated by a distance $L$.

Our scheme shares some similarities with the so-called hybrid quantum repeater such as the usage of hybrid entangled states and the dependencies and trade-off
related to the entanglement generation rate and state quality with regard to $\alpha\sin(\theta)$ where $\alpha$ is the optical coherent-state amplitude and $\theta$
is the angle of a spin-controlled phase rotation of the optical mode due to a dispersive light-matter-interaction.
However, due to the photonic middle stations in each repeater segment, our version inherits the twin-field-like scaling advantage. In some distance regimes this difference results in rates for our scheme that are larger by orders of magnitude compared with the original HQR version based on unambiguous state discrimination. For this version, we explicitly showed that the relevant quantities do not exhibit the twin-field-type square-root-enhancement of the transmission parameter per repeater segment like in our scheme.
We further showed that it is possible, in principle, to employ small dispersive phase rotations $\theta$ corresponding to weak optical non-linearities at the expense of using larger coherent state amplitudes $\alpha$ and more demanding phase stabilization.
Another advantage of our scheme compared to the original hybrid quantum repeater is that it is no longer necessary to couple nonclassical light states with a spin system (like an atom in a cavity) "inline".
It is now sufficient to prepare hybrid light-spin entangled states "offline" and couple the optical parts with beam splitters when executing the repeater protocol.

 For the $n=2$-case with only one memory station based on a parallel distribution scheme, we considered the most important imperfections like photon loss, detector inefficiencies, memory dephasing, dark counts, phase mismatch and faulty Bell measurements on the memories modeled by depolarization.
 This error analysis can also be extended to $n$ repeater segments when using a sequential distribution and swapping strategy. This approach enabled us to calculate exact BB84 secret-key rates (in the asymptotic limit) for the general case of $n$ repeater segments. For $n=2$ the parallel scheme outperforms the sequential one and for $n>2$ we have evidence that the sequential scheme is better. As we did not include quantum error correction, we focused on repeaters up to $n=16$.  
We calculated secret-key rates per channel use for realistic parameter regimes and showed that introducing a cut-off (maximal duration) for the memory waiting time can increase the secret-key rate enormously.

Our main quantitative results in terms of secret-key rates per channel use are that by introducing quantum memories
into a twin-field-based relay, for distances beyond 700 km,
the PLOB bound can be beaten with memory coherence times of 1s and modest detector
efficiencies. The ideal single-repeater scaling of $\sqrt{\eta_{total}}$
can be exceeded when coherence times of 5s and perfect detector efficiencies
are approached. In order to overcome the ideal twin-field rate only a coherence time of 1s is needed. Since our scheme is mainly for threshold detectors, but also involves
light-matter interactions, the light wavelengths must be suitably chosen
(possibly including additional frequency conversions which have not been
considered here) and the basic processing times, as usually in memory-based
quantum repeaters determined by classical communication times
and the speed of the light-matter operations, are longer than those
in twin-field QKD without memory assistance. Nonetheless, for sufficiently
many and short elementary segments, the scaling advantage of the memory-assisted
scheme can potentially overcome the disadvantage of the slower clock rates
(for phase-matching QKD without memories the source clock rate is just given by that of a laser
generating coherent states; creating cat states like in our BB84-type scheme is
unnecessary and so are light-matter couplings and classical waiting times).
We explicitly showed this by also considering secret-key rates per second.

We also investigated a variant of our scheme based on homodyne detectors.
According to our analysis, the regimes where a homodyne-based scheme
works is incompatible with the regimes where the scaling advantage
of a MA-PM QKD scheme becomes relevant.
Thus, secret-key rates for segments of 10 km and more are obtained to be zero
for the homodyne-based scheme. This is conceptually similar to the original
hybrid quantum repeater based on homodyne measurements where the
segment lengths also needed to remain sufficiently short (at around 10 km). A difference there,
however, was that additional quantum error detection (entanglement purification)
was included such that high-fidelity entangled states were still obtainable.
In our scheme, active methods for quantum error correction or detection
were not considered.

Like in all twin-field-type QKD approaches based on single-photon interference
or, more generally, interference of phase-sensitive single-mode states,
as opposed to those schemes relying on two-photon interference,
a means for robust phase stabilization must be included.
In our scheme, this could be achieved by sending a coherent-state
reference pulse along the fiber channels together with the signal pulses.

\begin{acknowledgments}
We thank the BMBF in Germany for support via Q.Link.X and the BMBF/EU for support via QuantERA/ShoQC.
We also thank Stefano Pirandola and Mark Wilde for useful comments on the manuscript.
\end{acknowledgments}
\appendix
\section{Error models}
\label{app:model}
Here we briefly describe all error models employed for our analysis.
A lossy channel with transmittance $\eta$  can be described as a beam splitter acting on the optical mode of interest $a$ and an environmental mode $b$ corresponding to the mode operator transformation

		\begin{equation}
			\begin{pmatrix}
				\hat{a}'\\
				\hat{b}'
			\end{pmatrix}
			=
			\begin{pmatrix}
				\sqrt{\eta}&\sqrt{1-\eta}\\
				\sqrt{1-\eta}&-\sqrt{\eta}
			\end{pmatrix}
			\begin{pmatrix}
				\hat{a}\\
				\hat{b}
			\end{pmatrix}	\,,			
		\end{equation}
where $\hat{a}'$ is the relevant output mode operator of interest and we trace out the environmental mode expressed by mode operator $\hat{b}'$. 
For fiber transmission, $\eta$ is given by $\exp(-\frac{L}{L_\text{att}})$, where $L$ is the fiber's length and $L_\text{att}$ is the attenuation length of 22km in a typical optical fiber. 

The dephasing of the memories is described by the following dephasing channel,

\begin{align}
\mathcal{E}_{dephasing}(t,T,\rho)=\nonumber\\\frac{1}{2}\left(1+\exp(-\frac{t}{T})\right)\rho+\frac{1}{2}\left(1-\exp(-\frac{t}{T})\right)Z \rho Z\,,\label{eq:dephasing_def}
\end{align}
where $\rho$ is a single-qubit density matrix, $Z$ the Pauli qubit phase-flip operator,  $t$ is the time for which the memory dephases and  $T$ is the memory coherence time.
The imperfections of the Bell measurement on the quantum memories is modeled by the following depolarizing channel,
\begin{align}
\mathcal{E}_{depol}(p_{depol},\rho)=\left(1-p_{depol}\right) \rho+  p_{depol} \frac{\mathds{1}}{2}\,.
\end{align}

The POVM element corresponding to a click of the on/off detector is given by
\begin{align}
\hat{E}= \mathds{1}-D(0)\ket{0}\bra{0}\,,\label{eq:darkcount_povm}
\end{align}
where $D(0)$ denotes the probability that the detector does not click on a vacuum state. This means the dark count probability is given by $1-D(0)$.
Fortunately, we will not require an explicit expression for the conditional density operator that incorporates dark counts, because we trace out the measured mode (see App. \ref{app:calc}).

\section{Approximation of $\mathds{E}[\max(X_1,\cdots,X_n)]$}
\label{app:approx}
In order to distribute entanglement over the whole distance of the repeater, entanglement needs to be generated in all $n$ segments.
When generating entanglement in the $n$ segments independently, the total waiting time is given by $\max(X_1,\cdots,X_n)$, where the geometrically distributed random variables $X_j$ describe the number of entanglement generation attempts until success in segment $j$ and where $p$ is the probability of success in a single attempt.
 Therefore, the raw rate scales inversely with $\mathds{E}[\max(X_1,\cdots,X_n)]$. This expectation value will also appear when we will discuss the dephasing in a parallel scheme using Jensen's inequality.
For the case $p\ll1$ and deterministic entanglement swapping it is possible to obtain a simple approximation of $\mathds{E}[\max(X_1,\cdots,X_n)]$ where X is geometrically distributed:
\begin{align}
\mathds{E}[\max(X_1,\cdots,X_n)]&=\sum_{j=1}^n \binom{n}{j} \frac{(-1)^{j+1}}{1-(1-p)^j}\\ &\approx \sum_{j=1}^n \binom{n}{j} \frac{(-1)^{j+1}}{j p}\,.
\end{align}
This approximation is based on the exact expression of Ref. \cite{hybridrepeaterrateanalysis} for arbitrary $p$.
We then expanded $\left(1-p\right)^j$ with the binomial theorem and neglected quadratic and higher orders of $p$.
We can furthermore prove by induction
\begin{equation}
\sum_{j=1}^n  \binom{n}{j}  \frac{(-1)^{j+1}}{j} =\sum_{j=1}^n \frac{1}{j}=:H(n)\,,
\end{equation}
where $H(n)$ are also known as harmonic numbers.
We  approximate the harmonic numbers by using only the first terms of their asymptotic expansion,
\begin{equation}
 H(n)\approx \gamma+ \ln(n)+\frac{1}{2n} \,,
\end{equation}
where $\gamma=0.57721\dots$ is the Euler-Mascheroni constant.
In the end we obtain the simple approximation
\begin{equation}
\mathds{E}[\max(X_1,\cdots,X_n)]\approx \frac{1}{p} \left( \gamma+\ln(n)+\frac{1}{2n}\right)\,.
\end{equation}
Note that this approximation scales with $\ln(n)$, while the widely used approximation $\left(\frac{3}{2}\right)^{\log_2(n)}\frac{1}{p}$ scales with $n^{\log_2(1.5)}$.
However, note that the latter depends on the assumption of both small $p$ and small swapping probabilities, so it is inapplicable here for deterministic swapping \cite{onthewaitingtime}.

\section{Effect of memory dephasing for $n=2$}
\label{app:dephasing}
For the case of two quantum repeater segments, the definition of $M_{par} \equiv M$ in Eq. (\ref{eq:par_dephasing_short}) simplifies to $|X_1-X_2|$ where $X_1$ and $X_2$ are independent geometrically distributed random variables.
Therefore, we have for the corresponding distribution
\begin{align*}
 \mathds{P}(M=0)=\sum_{k=1}^\infty \mathds{P}(X_1=X_2=k)=\sum_{k=1}^\infty p^2q^{2(k-1)}=\frac{p}{2-p} \,,
\end{align*}
 and for $j>0$,
\begin{align*}
&\mathds{P}(M=j)=\sum_{k=1}^\infty2 p^2q^{2(k-1)+j}=2\frac{p q^j}{2-p}\,,
\end{align*}
where the factor 2 comes from the fact that the two cases $X_1>X_2$ and $X_2>X_1$ are possible.

This allows us to calculate for $M:=|X_1-X_2|$
\begin{align}
\mathds{E}\left[\exp(-M \frac{\tau}{T})\right]=\frac{p}{2-p}\left(\frac{2}{1-q\exp(-\frac{\tau}{T})}-1\right)\,,
\end{align} 
and by summing only up to a constant instead of infinity and considering a renormalization, one can easily obtain the expectation value for protocols which abort after the memory has dephased for a given time (cut-off). The additional complexity of this protocol lies solely in the raw rate, which is already known in the literature \cite{rawratecutoffCollinsPRL,onthewaitingtime,Praxmeyer_finite_memory}.
Note that we also have to consider an additional non-random dephasing time, because each memory already dephased during the time between sending the optical mode and obtaining the information on whether the optical measurement was successful or not. Therefore, each memory dephases for a time unit of $\frac{L}{n c}$. If we perform the measurements on the two outer memories immediately, we only accumulate a constant dephasing time of $2(n-1)\frac{L}{n c}=\frac{2L}{c}(1-\frac{1}{n})$. If we perform the measurements of the outer memories at the end of the entanglement distribution (like in Eq. (\ref{eq:par_dephasing_long})), we accumulate a constant  dephasing time of  $\frac{2L}{c}$.

\section{Pauli channels and entanglement swapping}
\label{app:pauli}
We call a single-qubit channel $\mathcal{N}(\cdot)$ a Pauli channel if and only if $\mathcal{N}(\rho)=\sum_i p_i P_i\rho P_i^\dagger$ where $p_i$ are probabilities and $P_i$ are Pauli operators ($\mathds{1},X,Y,Z$).
Since all of these Pauli operators either commute or anti-commute, Pauli channels commute. The composition of two Pauli channels is again a Pauli channel, because the product of two Pauli operators is again a Pauli operator up to a phase which becomes irrelevant for the case of a Pauli channel since $P_i$ and $P_i^\dagger$ are both applied such that these phases cancel.  Since one can switch between all four two-qubit Bell states by applying one of the four single-qubit Pauli operators, it can be seen that every Bell-diagonal state is equivalent to a Pauli channel acting on a perfect Bell state.
Let us now show that Pauli channels commute with the entanglement swapping operation on perfect Bell states.

Without loss of generality we assume that the Bell measurement on two memory qubits for entanglement swapping yields $\ket{\Phi^+}$ as the measurement outcome, while the other three cases work analogously.
It is also sufficient to consider only two two-qubit pairs initially prepared in the Bell-states $\ket{\Phi^+}_{12}$ and  $\ket{\Phi^+}_{34}$ and each being partially subject to an arbitrary Bell-diagonal channel, $\mathcal{N}_2$ and $\mathcal{N}_3'$ for qubits 2 and 3
\begin{widetext}
\begin{align}
&\bra{\Phi^+}_{23}  \mathcal{N}_2(\ket{\Phi^+}_{12}\bra{\Phi^+})\otimes \mathcal{N'}_3(\ket{\Phi^+}_{34}\bra{\Phi^+}) \ket{\Phi^+}_{23}\nonumber \\
&=\bra{\Phi^+}_{23}\sum_{i,j=1}^4 p_i p_j'P_{i,2}\ket{\Phi^+}_{12}\bra{\Phi^+} P_{i,2}^\dagger \otimes P_{j,3}\ket{\Phi^+}_{34}\bra{\Phi^+} P_{j,3}^\dagger\ket{\Phi^+}_{23}\nonumber \\
&=\sum_{i,j=1}^4 p_i p_j'P_{i,1}P_{j,4} \bra{\Phi^+}_{23}\ket{\Phi^+}_{12}\bra{\Phi^+}  \otimes \ket{\Phi^+}_{34}\bra{\Phi^+}\ket{\Phi^+}_{23}P_{i,1}^\dagger P_{j,4}^\dagger\nonumber \\
&=\frac{1}{4}\sum_{i,j=1}^4 p_i p_j'P_{i,1}P_{j,4}\ket{\Phi^+}_{14}\bra{\Phi^+}P_{j,4}^\dagger P_{i,1}^\dagger \nonumber \\
&=\frac{1}{4}\sum_{i,j=1}^4 p_i p_j'P_{i,1}P_{j,1}\ket{\Phi^+}_{14}\bra{\Phi^+}P_{j,1}^\dagger P_{i,1}^\dagger \nonumber\\
&=\frac{1}{4}\mathcal{N}_1\left(\mathcal{N}'_1\left(\ket{\Phi^+}_{14}\bra{\Phi^+}\right)\right)\,. \label{eq:pauli_proof_end}
\end{align}
\end{widetext}
Here we used the fact that $P_{i,1}P_{i,2} \ket{\Phi^+}_{12}= \ket{\Phi^+}_{12}$ holds for all Pauli operators $P_i$  and we also employed that (qubit) Pauli operators are Hermitian and unitary and therefore self-inverse.

We can then apply this result for all entanglement swapping operations successively. 
Note that this argument relies on the assumption of Pauli channels/Bell-diagonal states, but initially when including detector dark counts the memory states are no longer Bell-diagonal and already dephasing before we apply a operation which erases the Bell non-diagonal elements  \cite[Sec. 3.2.1]{purification_review}. However, this erasing is done by applying random correlated two-qubit Pauli operations and hence commutes with the decoherence channel. As a consequence, we can first apply the erasing channel and therefore we have Bell-diagonal states (which are equivalent to a Pauli channel on a perfect Bell state) allowing us to use the result above.
 There is no additional temporal overhead due to the communication time needed for generating the correlations.
  For example, a memory could generate two correlated random variables and send one of them to the other memory belonging to this segment. The necessary communication time is given by $\frac{L}{n c}$, which is the same time as between sending the optical mode and obtaining the information whether the optical measurement succeeded or failed. Alternatively, the middle station could also generate the correlated random variables and send them to the memories if the optical measurement was successful. Therefore, only the amount of sent information by the middle station increases and thus there are no temporal issues.
In the end, we have to consider a concatenation of $n$ dephasing channels, each with a random decoherence time which is equivalent to a single dephasing channel where the dephasing time is given by the sum of all the individual dephasing times, e.g. $t+t'$ (assuming the same coherence time for both memories) for $\mathcal{N}_1$ and $\mathcal{N}_1'$ in Eq.  (\ref{eq:pauli_proof_end}) for $t$ as defined in Eq. (\ref{eq:dephasing_def}).
Similarly, we can simplify the concatenation of the $n-1$ depolarizing channels with parameter $p_{depol}$, describing the probability of no depolarization,  into a depolarizing channel with $1-p_{depol}'=\left(1-p_{depol}\right)^{n-1}$. 
The concatenation of the Pauli channel corresponding to dark counts/measurements cannot be simplified as much as for the depolarizing or dephasing channel.
For the concatenation of a general single-qubit Pauli channel, 
\begin{equation}
\mathcal{N}(\rho)=p_1 \rho+p_2 Z \rho Z +p_3 X\rho X+p_4 Y \rho Y \,,
\end{equation}
we obtain the following recursive set of equations,
\begin{align}
	\begin{pmatrix}
		p_1^{(n+1)}\\
		p_2^{(n+1)}\\
		p_3^{(n+1)}\\
		p_4^{(n+1)}
	\end{pmatrix}
=
	\begin{pmatrix}
		p_1 & p_2 & p_3 & p_4 \\
		p_2 & p_1 & p_4 & p_3 \\
		p_3 & p_4 & p_1 & p_2 \\
		p_4 & p_3 & p_2 & p_1
	\end{pmatrix}
	\begin{pmatrix}
		p_1^{(n)}\\
		p_2^{(n)}\\
		p_3^{(n)}\\
		p_4^{(n)}
	\end{pmatrix}\,,
\end{align}
where $p_1^{(0)}=1$ and $p_2^{(0)}=p_3^{(0)}=p_4^{(0)}=0$. Therefore, we have
\begin{align}
	\begin{pmatrix}
		p_1^{(n)}\\
		p_2^{(n)}\\
		p_3^{(n)}\\
		p_4^{(n)}
	\end{pmatrix}
	=\begin{pmatrix}
		p_1 & p_2 & p_3 & p_4 \\
		p_2 & p_1 & p_4 & p_3 \\
		p_3 & p_4 & p_1 & p_2 \\
		p_4 & p_3 & p_2 & p_1
	\end{pmatrix}^n
	\begin{pmatrix}
		1\\
		0\\
		0\\
		0
	\end{pmatrix}\,.
\end{align}
The transition matrix is real and symmetric and can thus be diagonalized, such that it is easy to calculate the power of the matrix.\\

\section{Calculation of the quantum repeater states with on/off detectors}
\label{app:calc}

Our simplest protocol ($n=1$) starts by creating hybrid entanglement at the two cavities (see Fig. \ref{fig:protocol} (b)), i.e. we first have the state 
\begin{align}
\frac{1}{2}(\ket{\uparrow,\uparrow,\alpha e^{-i\theta},\alpha e^{-i\theta}}+\ket{\downarrow,\downarrow,\alpha e^{i\theta},\alpha e^{i\theta}}\\ \nonumber
+\ket{\downarrow,\uparrow,\alpha e^{i\theta},\alpha e^{-i\theta}}+\ket{\uparrow,\downarrow,\alpha e^{-i\theta},\alpha e^{i\theta}}) \,.
\end{align}
After applying the lossy channels of transmittance  $\sqrt{\eta}$ (corresponding to the distance between Alice/Bob and the middle station) and the 50/50 beam splitter at the middle station we obtain the following state:
\begin{widetext}
\begin{align}
\frac{1}{2}\left(\ket{\uparrow,\uparrow,\sqrt{2\sqrt{\eta}}\alpha e^{-i\theta},0,\sqrt{1-\sqrt{\eta}}\alpha e^{-i \theta},\sqrt{1-\sqrt{\eta}}\alpha e^{-i \theta}}\right.\\ \nonumber
+\left. \ket{\downarrow,\downarrow,\sqrt{2\sqrt{\eta}}\alpha e^{i\theta},0,\sqrt{1-\sqrt{\eta}}\alpha e^{i \theta},\sqrt{1-\sqrt{\eta}}\alpha e^{i \theta}}\right.\\ \nonumber
+\left.\ket{\uparrow,\downarrow,\sqrt{2\sqrt{\eta}}\alpha\cos(\theta),-i\sqrt{2\sqrt{\eta}}\alpha \sin(\theta),\sqrt{1-\sqrt{\eta}}\alpha e^{-i \theta},\sqrt{1-\sqrt{\eta}}\alpha e^{i \theta}}\right.\\ \nonumber
+\left.\ket{\downarrow,\uparrow,\sqrt{2\sqrt{\eta}}\alpha\cos(\theta),i\sqrt{2\sqrt{\eta}}\alpha \sin(\theta),\sqrt{1-\sqrt{\eta}}\alpha e^{i \theta},\sqrt{1-\sqrt{\eta}}\alpha e^{-i \theta}}\right) \,.
\end{align}
\end{widetext}
Here, the last two entries in each ket vector represent the loss modes that initially start in a vacuum state.
In order to calculate the partial trace we will use the following calculation `trick'.
Suppose we are given a state of the form $\sum_k c_k \ket{k}_1\otimes \ket{\Psi_k}_2$ ($\ket{k}_1$ form an orthonormal basis, while $\ket{\Psi_k}_2$ may be arbitrary pure states) and we want to calculate the reduced density matrix of system 1:
\begin{align}
	\text{Tr}_2\left(\sum_{k,j} c_{k}c_j^* \ket{k}_1\bra{j}\otimes\ket{\Psi_k}_2\bra{\Psi_j}\right)\nonumber \\
=\sum_{k,j} c_{k}c_j^*  \text{Tr}_2\left(\ket{k}_1\bra{j}\otimes\ket{\Psi_k}_2\bra{\Psi_j}\right)\nonumber \\
=\sum_{k,j} c_{k}c_j^*  \ket{k}_1\bra{j}\sum_{l} \bra{l}_2 \ket{\Psi_k}_2\bra{\Psi_j}\ket{l}_2\nonumber \\
=\sum_{k,j} c_{k}c_j^*  \ket{k}_1\bra{j}\sum_{l} \bra{\Psi_j}\ket{l}_2 \bra{l}_2 \ket{\Psi_k}_2\nonumber \\
=\sum_{k,j} c_{k}c_j^*  \ket{k}_1\bra{j} \bra{\Psi_j}\ket{\Psi_k}_2 \label{eq:tracetrick}\,.
\end{align}
Similarly, one can show for the conditional state of subsystem 1 with measurement operators $A$ acting on subsystem 2:
\begin{align}
\text{Tr}_2\left(\sum_{k,j} c_{k}c_j^* \ket{k}_1\bra{j}\otimes A_2 \ket{\Psi_k}_2\bra{\Psi_j} A_2^\dagger\right)\nonumber \\
=\sum_{k,j} c_{k}c_j^*  \ket{k}_1\bra{j} \bra{\Psi_j}A_2^\dagger A_2\ket{\Psi_k}_2\,.
\end{align}
Note that $A^\dagger A$ is a POVM element and the POVM of an on/off detector including dark counts, see $\hat{E}$ of Eq. (\ref{eq:darkcount_povm}), is known in the literature \cite{kok_lovett_2010} and therefore we do not need to explicitly calculate a corresponding measurement operator $A$.
Moreover, there is no need to explicitly compute the effect of dark counts on the conditional states.
 This allows us to express all coefficients of the two memories' final density operator in terms of scalar products between coherent states.

If we measure the photon number (without dark counts) on the second optical mode after the beam splitter at the middle station, and trace out all other modes, we obtain the following density operator for Alice's and Bob's qubits:
\begin{widetext}
\begin{align}
\frac{1}{2}\left(\ket{\uparrow,\downarrow}\bra{\uparrow,\downarrow}+\ket{\downarrow,\uparrow}\bra{\downarrow,\uparrow}
\pm \bigl|\bra{\sqrt{1-\sqrt{\eta}}\alpha e^{i\theta}}\ket{\sqrt{1-\sqrt{\eta}}\alpha e^{-i\theta}}\bigr|^2(\ket{\uparrow,\downarrow}\bra{\downarrow,\uparrow}+\ket{\downarrow,\uparrow}\bra{\uparrow,\downarrow})\right)\,,
\end{align}
\end{widetext}
where $ |\bra{\sqrt{1-\sqrt{\eta}}\alpha e^{i\theta}}\ket{\sqrt{1-\sqrt{\eta}}\alpha e^{-i\theta}}|^2$ evaluates to $\exp\left[-4(1-\sqrt{\eta})\alpha^2\sin[2](\theta)\right]$. 
When considering only on/off detectors, the off-diagonal terms change and one additionally needs to take into account a factor of $\frac{\bra{-i\sqrt{2\sqrt{\eta}}\alpha \sin(\theta)}\left(\mathds{1}-\ket{0}\bra{0}\right) \ket{i\sqrt{2\sqrt{\eta}}\alpha \sin(\theta)}}{e^{2\sqrt{\eta}\alpha^2\sin[2](\theta)}-1}$ which simplifies to $-e^{-2\sqrt{\eta}\alpha^2\sin[2](\theta)}$. Therefore we obtain in total $e^{-2(2-\sqrt{\eta})\alpha^2\sin[2](\theta)}$ as the factor of the off-diagonal terms. 
This state is a mixture of two Bell states and, for the cases $n>1$, if we perform (ideal) Bell measurements on all $n$ segments, it is easy to see (due to the Pauli channel argument) that the exponent of the off-diagonal terms in the remaining state (after applying Pauli operations depending on the Bell measurement outcomes) is simply multiplied by $n$. 
For Bell-diagonal states with only two non-zero coefficients it is trivial to check that the distillable entanglement with only one-way classical communication coincides with the asymptotic secret-key fraction of BB84.

When considering also dark counts for the on/off detectors, we obtain the following (unnormalized) state:
\newline
\FloatBarrier
\begin{table}[h]
\begin{tabular}{l|l|l|l|l}
                              & $\bra{\uparrow,\uparrow}$ & $\bra{\downarrow,\downarrow}$ & $\bra{\uparrow,\downarrow}$ & $\bra{\downarrow,\uparrow}$ \\ \hline
$\ket{\uparrow,\uparrow}$     & $a$                       & $c^*$                         & $d_1^*$                       & $d_2^*$                       \\ \hline
$\ket{\downarrow,\downarrow}$ & $c$                       & $a$                           & $d_2$                         & $d_1$                         \\ \hline
$\ket{\uparrow,\downarrow}$   & $d_1$                       & $d_2^*$                         & $b$                         & $f^*$                       \\ \hline
$\ket{\downarrow,\uparrow}$   & $d_2$                       & $d_1^*$                         & $f$                         & $b$                        
\end{tabular}
\end{table}
\FloatBarrier
 with $a=\bra{0}\hat{E}\ket{0}=1-D(0)$ where $\hat{E}$ is the click operator considering dark counts \cite{kok_lovett_2010}, and D(0) is the probability that the detector does not click when a vacuum state is used as the input. Further, we have
\begin{align}
b&=\bra{\pm i \sqrt{2\sqrt{\eta}}\alpha \sin(\theta)}\hat{E}\ket{\pm i \sqrt{2\sqrt{\eta}}\alpha \sin(\theta)}\nonumber\\&=1-e^{-2\sqrt{\eta}\alpha^2\sin[2](\theta)}D(0)\,,
\end{align}
\begin{widetext}
\begin{align}
c&=\bra{\sqrt{1-\sqrt{\eta}}\alpha e^{-i \theta}}\ket{\sqrt{1-\sqrt{\eta}}\alpha e^{i \theta}}^2\bra{\sqrt{2\sqrt{\eta}}\alpha e^{-i \theta}}\ket{\sqrt{2\sqrt{\eta}}\alpha e^{i \theta}} a \nonumber\\ &=e^{2\alpha^2(\exp(2i \theta)-1)} a=a  e^{-4\alpha^2\sin[2](\theta)+i 2\alpha^2 \sin(2\theta)}\,,\\
d&=d_1=d_2=\bra{\sqrt{1-\sqrt{\eta}}\alpha e^{-i \theta}}\ket{\sqrt{1-\sqrt{\eta}}\alpha e^{i \theta}}\bra{\sqrt{2\sqrt{\eta}}\alpha\cos(\theta)}\ket{\sqrt{2\sqrt{\eta}}\alpha e^{i \theta}}\bra{0}\hat{E}\ket{i \sqrt{2\sqrt{\eta}}\alpha \sin(\theta)}\nonumber\\&=a  e^{-2\alpha^2 \sin[2](\theta)+i \alpha^2 \sin(2\theta)}\,,\\
f&=|\bra{\sqrt{1-\sqrt{\eta}}\alpha e^{-i \theta}}\ket{\sqrt{1-\sqrt{\eta}}\alpha e^{-i \theta}}|^2\bra{i\sqrt{2\sqrt{\eta}}\alpha \sin(\theta)}\hat{E}\ket{-i\sqrt{2\sqrt{\eta}}\alpha \sin(\theta)}\nonumber\\&=e^{-2\alpha^2\sin[2](\theta)(2-\sqrt{\eta})}(e^{-2\sqrt{\eta}\alpha^2\sin[2](\theta)}-D(0))\,.
\end{align}
\end{widetext}
Note that without dark counts, $a=c=d=0$ and $D(0)=1$, we recover the effective $2\times 2$ matrix of the loss-only case.
A distinction between $d_1$ and $d_2$ has to be made when we consider entanglement swapping strategies which do not double the distance.\\
 Note that the phases of these parameters now also have a $\alpha^2 \sin(2\theta)$ dependency while there was no such dependency in the ideal case without dark counts. If we transform the state into a Bell-diagonal state we have the parameter $c$ which gives use information about the relative distribution of $\ket{\phi^\pm}$ and this parameter varies periodically with $\theta$. Therefore, it can be useful to apply local transformations for permuting the four Bell-state coefficients \cite{localbellpermutation} in order to obtain a higher secret-key fraction using BB84. 
When considering a swapping scheme where entanglement swapping is performed between two segments of equal size, one obtains the following set of recursive equations describing the unnormalized two-qubit state (assuming $2^j$ elementary segments and $\ket{\Phi^+}$ as measurement outcome, while above we considered the case of $j=0$ and omitted the subscript):
\begin{align}
a_{j+1}&=a_{j}^2+b_{j}^2+2 \text{Re}(d_{j}^2)\,,\nonumber\\
b_{j+1}&=2(a_{j} b_{j}+\text{Re}(d_{j}^2))\,,\nonumber\\
c_{j+1}&=2d_{j}^2+f_{j}^2+c_{j}^{* 2}\,,\\
d_{j+1}&=d_{j}(a_{j}+b_{j}+c_{j}^*)+d_{j}^*f_{j}\,,\nonumber\\
f_{j+1}&=2(|d_{j}|^2+f_{j} \text{Re}(c_{j}))\,.\nonumber
\end{align}
Note that for $n=1$ the BB84 secret-key fraction is not reduced due to discarding the off-diagonal terms in the Bell basis.
For $n=2$, the effect of discarding them is negligibly small. Also note that the approach here that leads to these recursive equations does not yield the same rates as using the protocol version based on the results of Ref. \cite{teleportation_as_depol} without correlated Pauli operations (see  App. \ref{app:discuss}.2), because we do not average over all possible Bell measurement outcomes. The calculation of the reduced state considering phase mismatch is completely analogous.
\section{Errors beyond loss, homodyne detection}
\label{app:discuss}
\subsection{Memory dephasing}
Let us consider $n$ repeater segments ($n>1$, otherwise no memory is needed). We can then assign independent random variables $X_j$ ($j\in\{1,\cdots,n\}$) to every segment counting for each the number of attempts until the entanglement is distributed due to a successful measurement outcome of the detector(s) for that segment.
These random variables follow a geometric distribution $\mathds{P}(X=k)=p q^{k-1}$ with $q=1-p$ where $p$ is the probability for a successful measurement outcome.
We can then introduce a new random variable $M$, which is a function of the random variables $X_j$,  describing the totally accumulated memory time for which the quantum states dephase.
Note that the specific form of the random variable $M$ differs for different entanglement generation/swapping protocols. In Sec.\ref{sec:quantumrepeater}.B we only considered a scheme where entanglement distributions in the $n$ segments are done in parallel. In terms of the raw rate it is clear that such a scheme achieves better rates than any sequential approach. However, when we also consider finite memory times it is no longer obvious whether the parallel scheme still performs better in terms of secret-key rate, because it is possible during the parallel distributions that multiple segments dephase simultaneously resulting in a longer accumulated memory dephasing time. In contrast, in an appropriate sequential scheme where always only one pair is distributed and swapping is immediately performed as soon as two pairs are present next to each other, at most a single memory pair is subject to a longer dephasing at any time.  

In the special case of $n=2$ it is impossible that multiple memory pairs dephase simultaneously and therefore in terms of secret-key rate the parallel scheme ($M=2|X_1-X_2|$) outperforms the sequential one ($M=2X_2$). The factor two here takes into account the situation when there are two memories dephasing in each segment. It is intuitive that for $n=2$ the parallel scheme outperforms the sequential one for two reasons. First, in the parallel scheme we only need to wait $\max(X_1,X_2)$ time steps instead of $X_1+X_2$ in order to distribute entanglement in both segments. Second, the memories also dephase to a lesser extent in the parallel scheme, because in both schemes at most one memory pair has to wait, but in the parallel scheme it is also possible that both segments succeed simultaneously. In general, for $n$ segments, the raw rate in a sequential scheme is given by $\frac{p}{n}$, while in a parallel scheme it is given by $\frac{p}{H(n)}$, where $H(n)$ is the $n$th harmonic number (assuming $p\ll1$, see App. \ref{app:approx}). Let us emphasize that this raw rate approximation holds for any memory-based quantum repeater that distributes entanglement in parallel and operates without nested quantum error detection/correction. However, for $n>2$ it is easy to calculate the average dephasing for the sequential scheme exactly while it is more complicated for the parallel one. In order to calculate it for a parallel scheme we assume that entanglement swapping is performed when entanglement was distributed in all segments in order to simplify the analysis (see App. \ref{app:sequential}). We found that the sequential scheme always gives better secret-key rates than our simple parallel scheme (except for $n=2$). This comparison is based on both exact and lower bounded dephasing factors for the sequential scheme together with lower bounds on the secret-key fraction for the parallel scheme. Supported by this, whenever memory dephasing is included, we shall consider the parallel scheme for the $n=2$ case and the sequential scheme otherwise ($n>2$). Thus, our focus on the sequential scheme for $n>2$ has two benefits: the secret-key rates can be calculated exactly and they turn out to be better thanks to the reduced total average dephasing. The inferior raw rates, $\frac{p}{n}$ versus $\frac{p}{H(n)}$ for the parallel scheme, appear to have a smaller impact on the secret-key rates (for up to $n=16$, the difference is a factor smaller than five).

The resulting random  state of a single protocol run with on/off detectors is then given by the density matrix:
\begin{align}
\label{eq:dephasedstate}
&\frac{1}{2}\left(1+ e^{-2n(2-\sqrt{\eta})\alpha^2\sin[2](\theta)} \exp\left(-M\frac{\tau}{T}\right)\right)\ket{\Phi^+}\bra{\Phi^+}\nonumber\\ \nonumber+&\frac{1}{2}\left(1- e^{-2n(2-\sqrt{\eta})\alpha^2\sin[2](\theta)} \exp\left(-M\frac{\tau}{T}\right)\right)\ket{\Phi^-}\bra{\Phi^-}\,,\\
\end{align} 
where $\tau$ is the duration of a single entanglement generation attempt in one segment and $T$ is the coherence time of the memory. 
Note that this state corresponds to the final state shared between Alice and Bob over the total channel distance (while for the case of Alice and Bob immediately measuring their qubits it is an effective rather than a physically occurring state).

The density operator in Eq. (\ref{eq:dephasedstate}) describes the state after a single run, but we are interested in the averaged state.
This means we have to calculate the expectation value $\mathds{E}[\exp(-M\frac{\tau}{T})]$. 
We calculate this expectation value for the case $n=2$ for the parallel scheme in App. \ref{app:dephasing}. In a sequential scheme the expectation value $\mathbb{E}[\exp\left(-M\frac{\tau}{T} \right)]$ can be calculated easily for arbitrary $n$, because $M$ is simply a sum of independent and identically distributed geometric random variables, whereas for a parallel scheme it is generally not known how to calculate the expectation value for arbitrary $n$. In App. \ref{app:sequential} we will discuss a lower bound on the secret-key rate based on Jensen's inequality when using a parallel scheme with arbitrary $n$.

Since we are here only interested in the secret-key rate we do not need to consider distributing physical entanglement over the whole distance. This means we can perform the measurement on Alice's and Bob's memories in the beginning with no need for waiting until the entanglement is distributed over the whole repeater. For the parallel scheme this has only a little effect by improving
\begin{align}
\label{eq:par_dephasing_long}
M_{\text{par}}=2\sum_{j=1}^{n}\left(\max(X_1,\dots,X_n)-X_j\right)
\end{align}
to 
\begin{align}
\label{eq:par_dephasing_short}
M_{\text{par}}=&2\sum_{j=2}^{n-1}\left(\max(X_1,\dots,X_n)-X_j\right)\\&+2\max(X_1,\dots,X_n)-X_1-X_n\,. \nonumber
\end{align} 
For Eq. (\ref{eq:par_dephasing_short}), in the segments next to Alice and Bob there is only one memory dephasing instead of two like for Eq. (\ref{eq:par_dephasing_long}).

In the case of the sequential scheme we can improve
\begin{align}
\label{eq:seq_dephasing_long}
M_\text{seq}=2\sum_{j=2}^{n}X_j
\end{align}
to
\begin{align}
\label{eq:seq_dephasing_short}
M_\text{seq}=\sum_{j=2}^{n}X_j\,,
\end{align}
since in the sequential scheme there is always a single segment dephasing where for Eq. (\ref{eq:seq_dephasing_short}) we removed the dephasing in one of the two memories. Therefore, we effectively double the memory coherence time for arbitrary $n$, whereas in the parallel scheme the improvement reduces with increasing segment number $n$.

Due to the finite memory time it is useful to consider a cut-off parameter which defines a maximal decoherence time before a state is discarded.
For the case of only two segments we have calculated the expectation value of the dephasing fractions with cut-off.
Here our focus is mainly on repeaters with $n=2,3,4$ repeater segments whose ultimate secret-key rates per channel use scale as $\sqrt[4]{\eta_{total}},\sqrt[6]{\eta_{total}}$, and $\sqrt[8]{\eta_{total}}$, respectively.
\begin{figure}
	\includegraphics[width=0.4\textwidth]{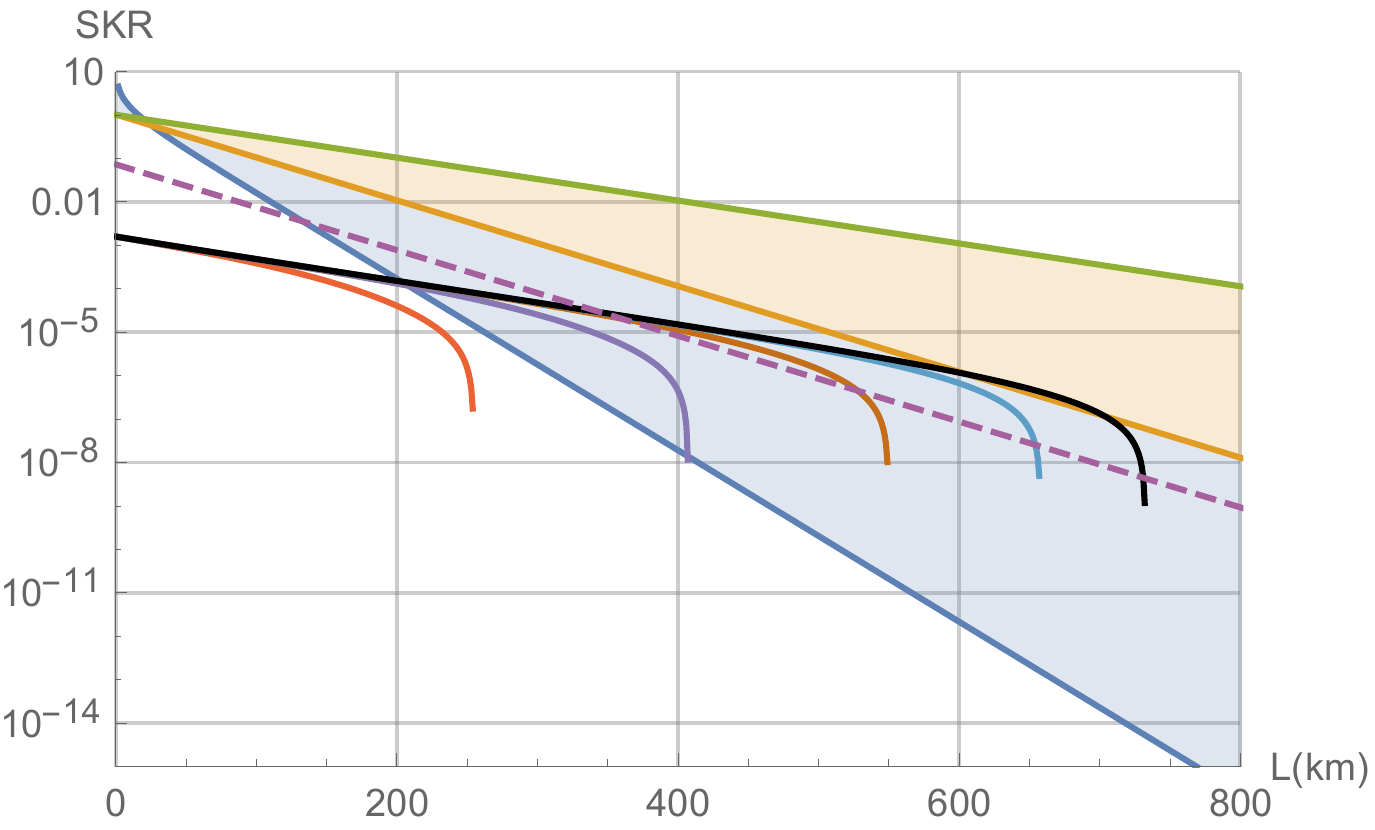}
	\caption{(Color online) Secret-key rates for a two-segment repeater ($n=2$, parallel scheme) without phase mismatch and assuming the parameters as listed in the main text. The straight lines (from bottom to top) denote the PLOB bound,$\sqrt{\eta_{total}}$, and $\sqrt[4]{\eta_{total}}$. The rates are for different coherence times $T$ of (1,10,100,1000,$\infty$) seconds (from left to right). The areas between PLOB and $\sqrt{\eta_{total}}$ and between $\sqrt{\eta_{total}}$ and $\sqrt[4]{\eta_{total}}$ are highlighted in color. The purple, dashed line denotes the loss-only case of standard twin-field QKD with perfect detector efficiencies and assuming a coherent-state amplitude optimized for the regime of large loss \cite{twinfield_luetkenhaus}.} 
	\label{fig:comparison_memory}
\end{figure}

\begin{figure}
	\includegraphics[width=0.4\textwidth]{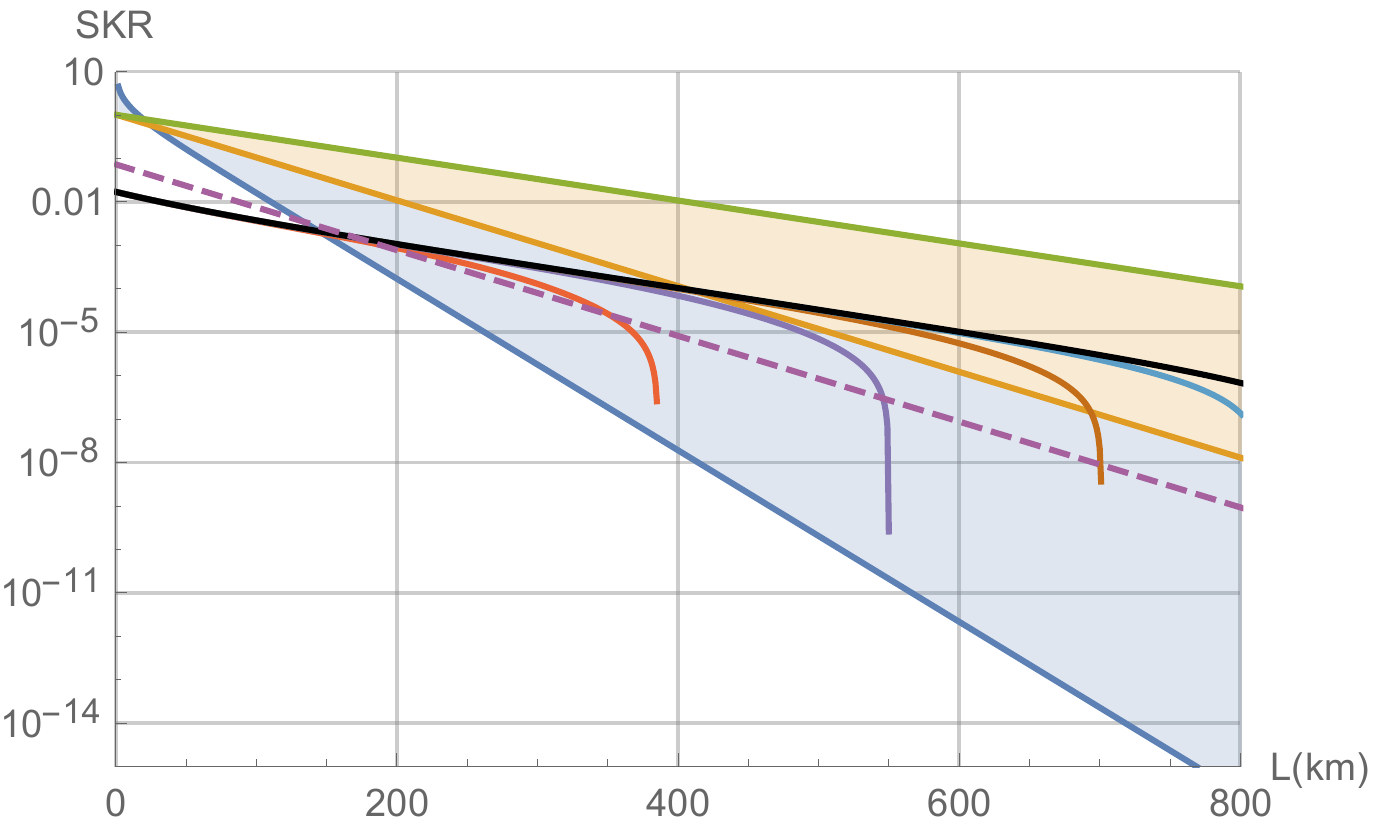}
	\caption{(Color online) Secret-key rates assuming the same parameters as in Fig. \ref{fig:comparison_memory} except for $p_{det}=1$ instead of $p_{det}=0.15$.
	}
	\label{fig:comparison_memory2}
\end{figure}
\subsection{Dark counts and phase mismatch}
With the inclusion of detector dark counts we need to use the full  $4\times4$ density matrix (in the computational basis) instead of an (effective) $2\times2$ matrix (in the case without dark counts all matrix elements except a $2\times2$ submatrix were zero) in order to describe the two-qubit state.
Calculating the state before the entanglement swapping  is straightforward but lengthy (see App. \ref{app:calc}) and the state after multiple entanglement swappings can be described by a set of recursive relations (see also App. \ref{app:calc}). 
In order to simplify the analysis we apply classically correlated Pauli operations to both parts of the imperfect Bell states, such that we erase the off-diagonal terms in the Bell basis \cite[Sec. 3.2.1]{purification_review}.
We do not need to let the memories dephase additionally for obtaining the classical correlations as required for the correlated Pauli operations, because an entanglement generation attempt takes $\tau=2\frac{L_0}{2c}$ in order to send the optical mode to the detector in the middle of the segment (length $L_0$) and to learn the measurement outcome.
If one party sends the bits for establishing classical correlations at the same time as it sends the mode to the detector, then we do not get an additional temporal overhead. 
As a consequence, this allows us to describe all errors as Pauli channels which act onto perfect Bell states.
Therefore, we can conduct our analysis as if we perform the entanglement swapping on perfect Bell states and apply all the errors afterwards (see App. \ref{app:pauli}).
Also notice that it is possible to obtain the advantage of a simplified analysis without the need for correlated Pauli operations \cite{teleportation_as_depol}.
In this case one performs entanglement swapping as usual, i.e. one applies Pauli corrections depending on the measured Bell state, but after the Pauli correction one discards the information  about the measurement outcome. Due to this averaging the teleportation reduces to a Pauli channel. Therefore, we can also interpret our protocol as applying $n-1$ teleportation steps (each represented by a Pauli channel) onto a non-Bell-diagonal state. Since a channel is linear, we can split the  non-Bell-diagonal state into a Bell-diagonal part and a part containing the off-diagonal elements. When applying the Pauli channel to these two parts, we see that the first part is exactly the state we considered in the previous protocol. In the second part the Bell states are simply permuted by Pauli operations, such that the state after applying the Pauli channels again only contains off-diagonal elements. However, these off-diagonal elements do not matter for the BB84 secret-key rate.
Note that these simplifications (applying correlated Pauli operations or discarding the measurement outcome) are at the expense of a worse secret-key rate in comparison to the case without correlated Pauli operations where we still keep track of the measurement outcome and do not average.

We compared the secret-key fraction of the simplification and the exact case (for $n=2$) using the parameters as mostly chosen in Sec. \ref{sec:comparison}. For this comparison, we considered loss and dark counts with parameters as in Sec. \ref{sec:comparison}. We found that the relative error increases exponentially with the distance of the total repeater. However, only for distances that are just a bit shorter than the distance where the secret-key fraction drops to zero the relative error becomes relevant, up to the point when the relative error diverges near the point where the secret-key fraction drops to zero. Therefore, we conclude that it is safe to use this simplification when not considering the neighborhood of the point where the secret-key fraction drops to zero.

In order to allow for phase mismatch errors which occur e.g. due to small differences in the laser frequencies and length fluctuations of the optical path, we model this error by assuming that one party employs a coherent state with amplitude $\alpha$ for generating the hybrid entangled states while the other party uses a coherent state with amplitude $\alpha e^{i \phi}$, where $\phi$ is a random variable with, for simplicity, a uniform distribution on the interval $(-\frac{\Delta}{2},\frac{\Delta}{2})$. We also have to bear in mind that this random phase difference has an influence on the raw rate (depending on $\alpha \sin(\theta)$) and especially for a small dispersive phase rotation $\theta$ the rate can vary up to a few percent. However, the relevant distribution for the secret-key fraction is the probability distribution of $\phi$ after conditioning onto a detector click. Therefore the relevant distribution is not uniform anymore, but larger values of $|\phi|$ have a larger probability (up to the point where the probability drops to zero).
Nevertheless, the difference between the actual and uniform distributions is small.
We calculated the Bell-diagonal coefficients and their expectation values with respect to $\phi$. However, even for the uniform distribution it is only possible to calculate the expectation value by numerical integration and therefore one could easily consider a more realistic model for the distribution of the phase difference $\phi$. 

\begin{figure}
	\subfloat[\label{sfig:phase_ideal_memory}]{
		\includegraphics[width=0.4\textwidth]{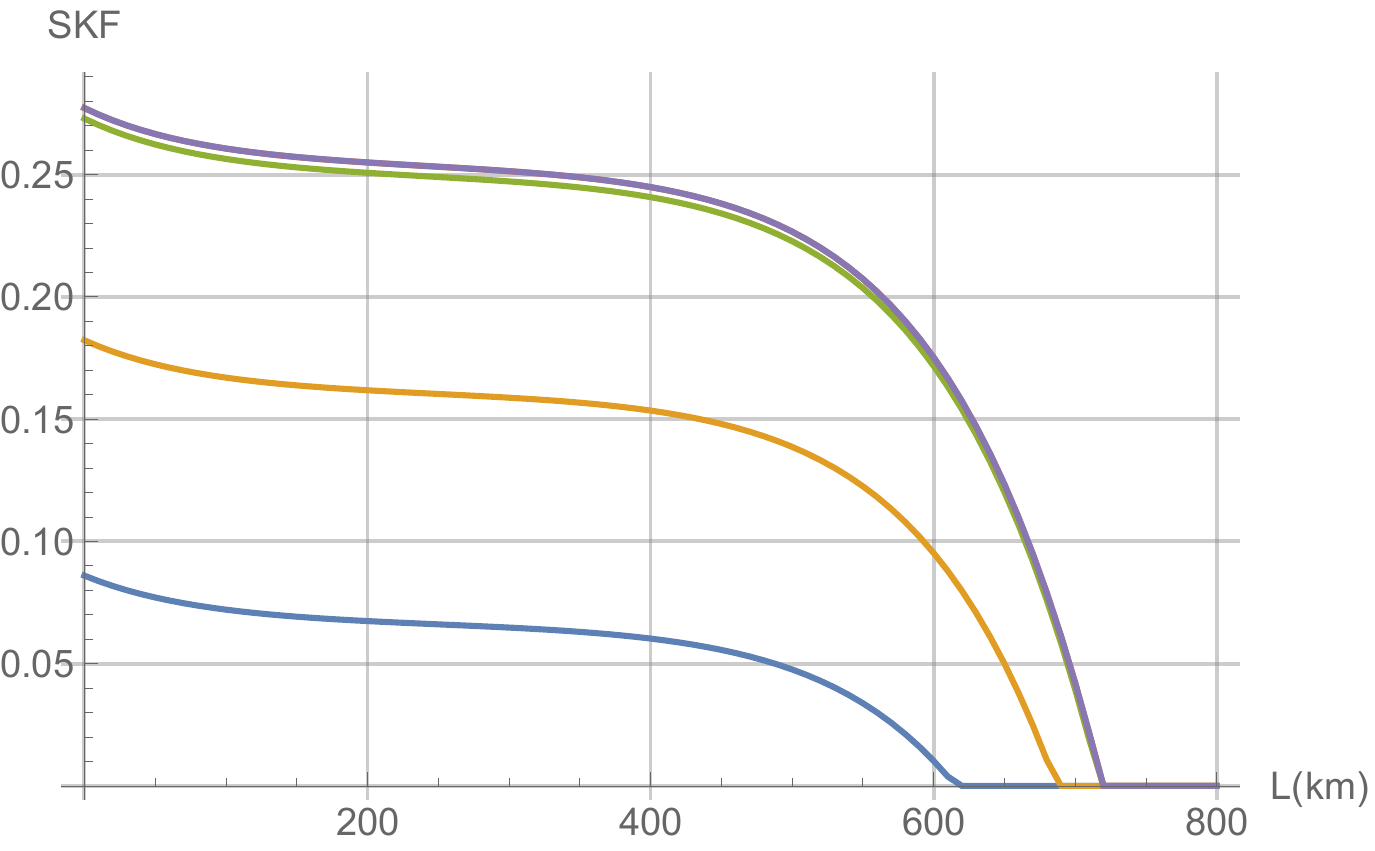}
	}\\
	\subfloat[\label{sfig:phase_10factor}]{
		\includegraphics[width=0.4\textwidth]{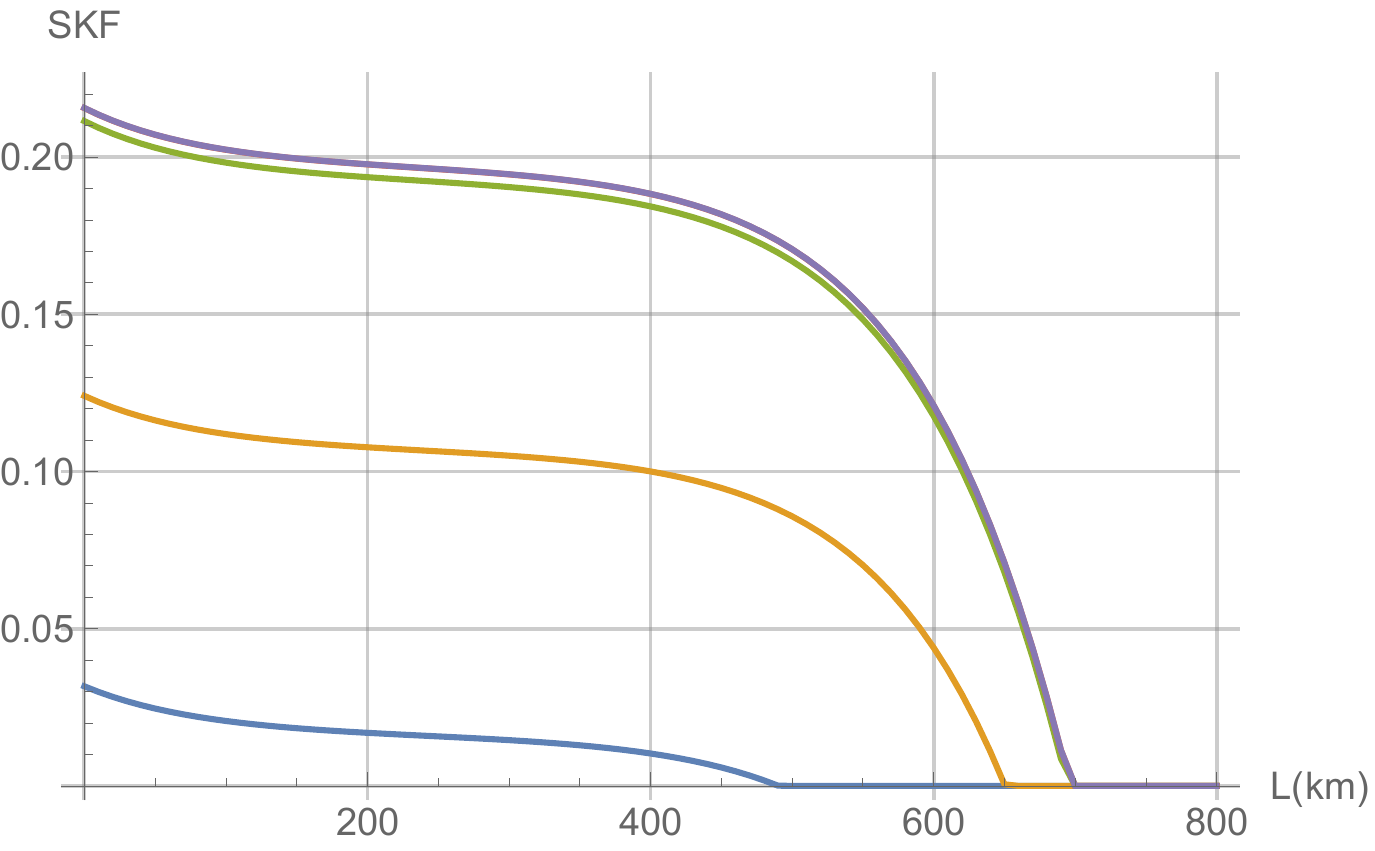}
	}\\
	\subfloat[\label{sfig:phase_1factor}]{
		\includegraphics[width=0.4\textwidth]{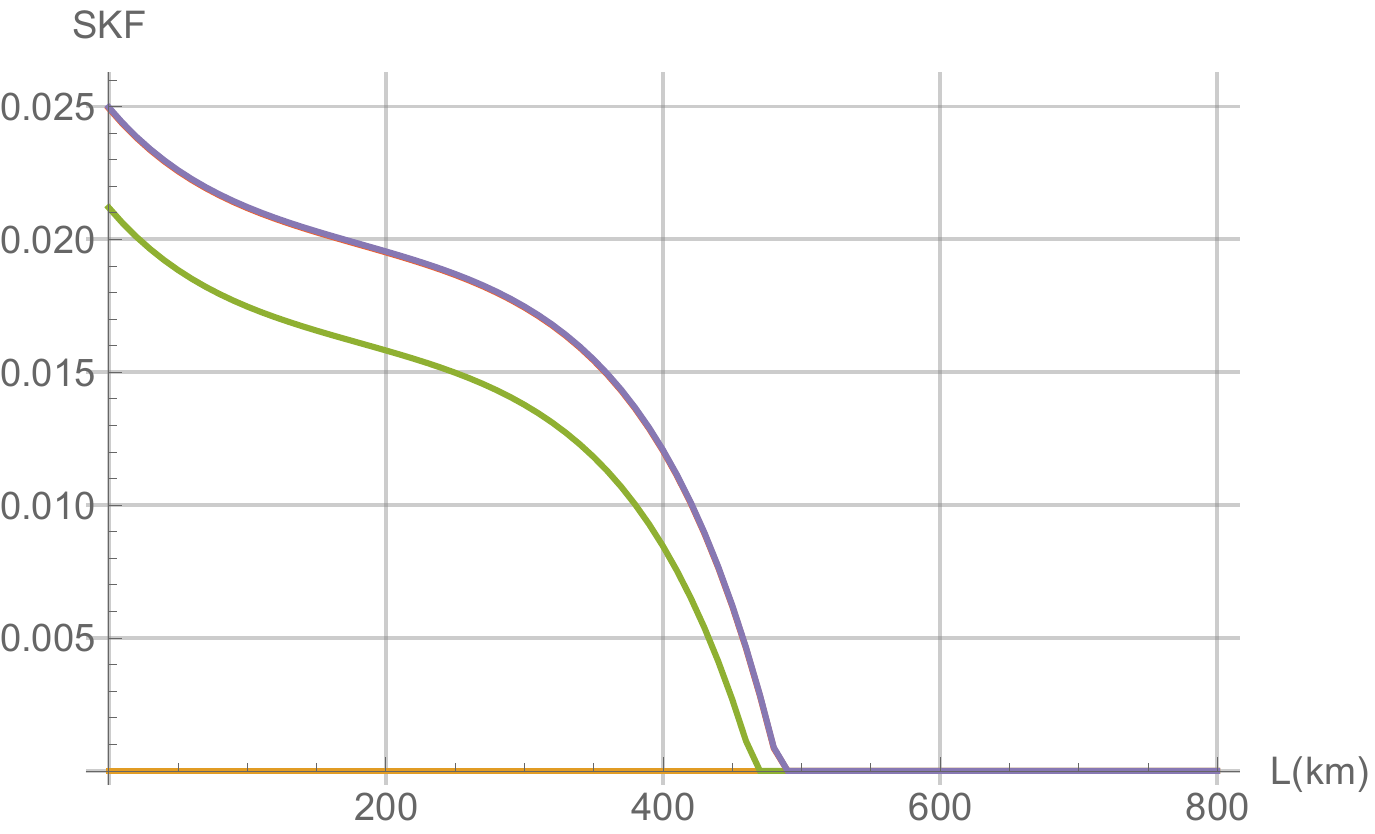}
	}
	\caption{(Color online) Secret-key fraction for the two-segment quantum repeater ($n=2$, parallel scheme) using the parameters discussed in Sec.\ref{sec:comparison} . We choose different memory coherence times for the three different plots and in each plot we consider a phase mismatch $\Delta$ of ($0, 10^{-4},10^{-3},5\times10^{-3},7.5\times 10^{-3}$) (from top to bottom).
		(a) ideal memories, (b) $T=10\mathds{E}(M)\tau$, (c)  $T=\mathds{E}(M)\tau$.}
	\label{fig:phase_comp}
\end{figure}

According to Fig. \ref{fig:phase_comp} the phase mismatch can be almost neglected when $\Delta<0.1\theta$ (this even holds for $\theta=\frac{\pi}{2}$).  However, for larger $\Delta$ the secret-key rate drops to zero very fast.
For $\Delta=\theta=0.01$ it is even impossible to obtain a secret key using the above parameters.
Therefore, we cannot choose $\theta$ arbitrarily small since this increases too much the required precision of the phase matching.

\subsection{Homodyne measurement}

In the main part of the paper, we only consider a scenario where Charlie (besides the less practical case of PNRDs) employs an on/off-detector. This is similar to previous twin-field QKD schemes.
However, it is straightforward to treat homodyne measurements for the two modes instead.
Homodyne measurements have the benefit of  near-unit efficiencies.
When reconsidering Eq. (\ref{eq:state}) one can see that the state shares some similarities to that of the HQR in Eq. (\ref{eq:hybridrepstate}).
If we can discriminate the peak at 0 from those at $\pm \sqrt{2}\alpha \sin(\theta)$ in the first mode  with a $p$-measurement (imaginary part of $\sqrt{2}\alpha \cos(\theta)$ versus that of $\sqrt{2}\alpha \exp(\pm i\theta)$ for, recall, $\alpha\in\mathbb{R}^+$), then we only learn that Alice and Bob have different bits but we do not learn their values.
However, in order to not learn their values by measuring the second mode (to disentangle it from the remaining system) we need to measure the $x$-quadrature in the second mode (real part of $\pm i \sqrt{2}\alpha\sin(\theta)$). 
It is also possible to exchange the two modes by which one obtains the same secret-key fraction after a suitable postselection of states.
The actual calculation is similar to that with on/off detectors and can be found in App. \ref{app:homodyne}.
Using homodyne measurements it is not obvious how to define a successful detector event.
We will consider an event to be successful if the measurement result of the quadrature $p_1$ lies within the interval $(-\Delta_p,\Delta_p)$, and the measurement result of $x_2$ must also occur within the interval $(-\Delta_x,\Delta_x)$. Choosing $\Delta_x$  and $\Delta_p$ is a compromise between a high raw rate and a high state quality.
For a given $\alpha$ and $\theta$ we can reduce the $Z$-error rate by decreasing $\Delta_p$. One might think that the parameter $\Delta_x$ is not relevant and can therefore be set to $\infty$.
However, this is not true since it also has an influence on the $X$-error rate making it even impossible to share a secret key in the no-loss case of $\sqrt{\eta}=1$ for too large $\Delta_x$.
This problem can be solved by simply choosing a sufficiently small $\Delta_x$, but even then a non-zero secret-key rate cannot be obtained for even moderate losses like $\sqrt{\eta}=0.7$ (about 8 km for the physical segment length assuming perfect detectors).

\section{Calculation of the quantum repeater states with asymmetric link lengths}
\label{sec:asym}

\begin{figure}
\includegraphics[width=0.45\textwidth]{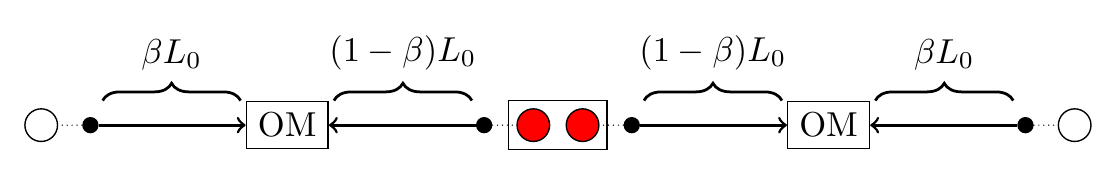}
\caption{(Color online) Asymmetric variation of our proposed scheme for $n=2$. The beam splitter is placed nearer to the memory station ($\beta>\frac{1}{2}$) such that the overall repetition rate can be increased. Note that for $n>2$ there is no gain with this variation. Due to this asymmetry Alice (as well as Bob) and the central memory station have to choose different amplitudes of the coherent states and we denote the amplitude arriving at the beam splitter by $\alpha_{BS}$.} 
\label{fig:protocolnequal2_asym}
\end{figure}
In this appendix, we discuss the obtainable secret-key rates per second in an asymmetric setting (for $n=2$) where the beam splitters are placed closer to the central memory station and farther away from Alice and Bob. This way, compared with the fully symmetric scheme, repetition rates can be increased (thanks to shorter classical communication times) at the expense of a worse scaling with distance.
Similar to the case with symmetric link lengths as discussed in App. \ref{app:calc}, we can calculate the resulting two-qubit state in the asymmetric setting as illustrated in Fig. \ref{fig:protocolnequal2_asym}. When we consider the loss-only case, we obtain
\begin{widetext}
\begin{gather}
a=c=d=0\,,\nonumber\\
b=1-\exp(-2\alpha_{BS}^2\sin[2](\theta))\,,\\ \nonumber
f=(\exp(-2\alpha^2_{BS}\sin[2](\theta))-1)\exp(-2\alpha_{BS}^2\sin[2](\theta))\\ \times \exp(-\alpha^2_{BS}\exp(\frac{L_0}{L_{att}})(\exp(-\frac{L_0 \beta}{L_{att}})(1-e^{-2i\theta})+\exp(-\frac{L_0(1-\beta)}{L_{att}})(1-e^{2i\theta})-2(1-\cos(2\theta))))\,.\nonumber
\end{gather}
\end{widetext}
Here, $\beta$ describes the asymmetry of the scheme as follows. The distance from Alice/Bob to the beam splitter is given by $\beta L_0$ and the distance between the memory and the beam splitter is therefore given by $(1-\beta)L_0$.
Since Alice/Bob have different distances to the beam splitter compared with the memory, both parties need to use different amplitudes in the light-spin entangled states. We choose their amplitudes in such a way that the amplitude at the beam splitter is given in both cases by $\alpha_{BS}$
 
\begin{figure}
\includegraphics[width=0.4\textwidth]{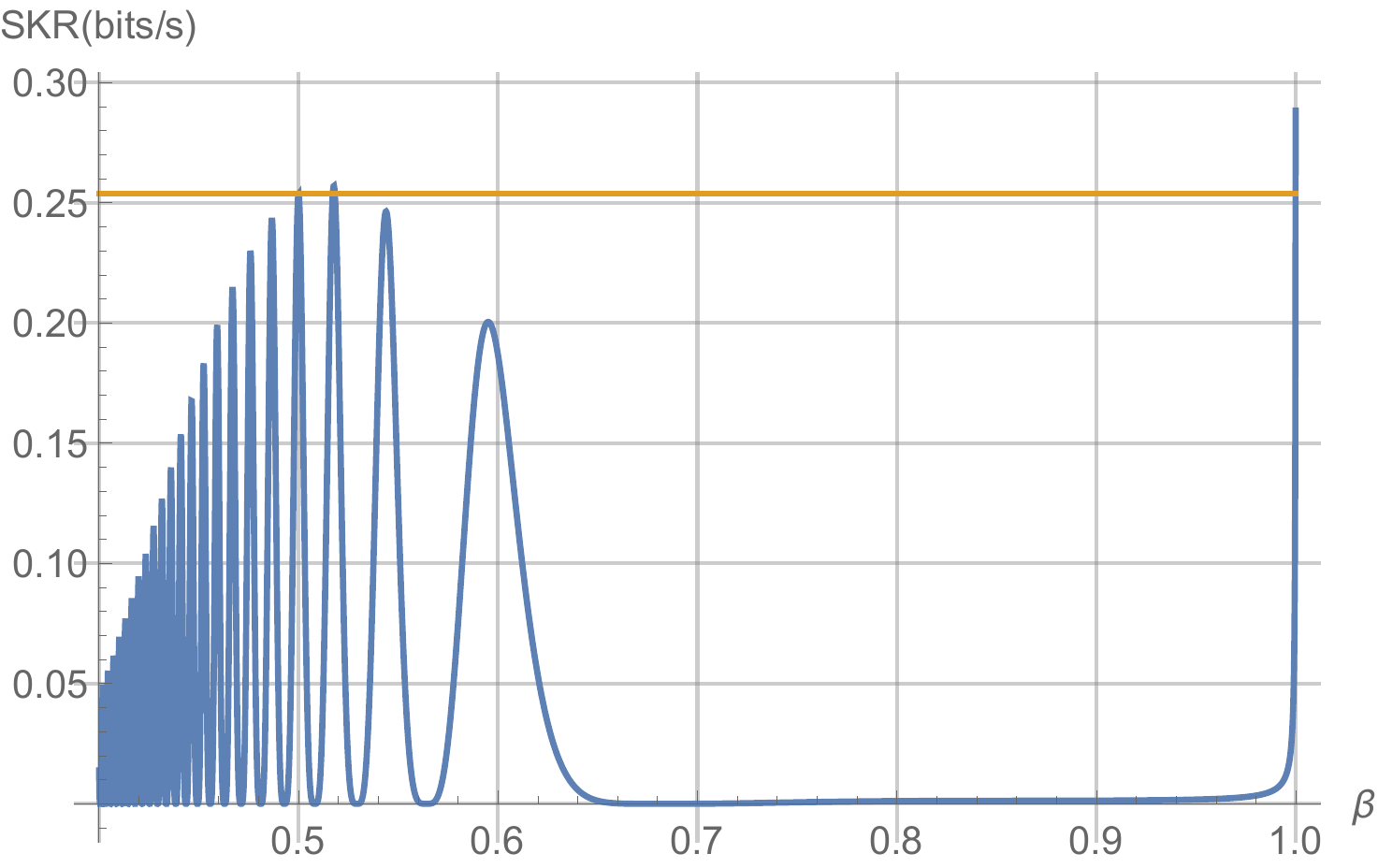}
\caption{(Color online)  Secret-key rate per second in the loss-only case of our asymmetric parallel scheme ($n=2$) for a total distance of 400 km in dependence of the asymmetry-parameter $\beta$. The constant line is given by the secret-key rate of the symmetric scheme ($\beta=\frac{1}{2}$). We assume that the repetition rate is limited to 10 MHz because of local operation times.  In this case, we can comfortably beat the symmetric scheme for strong asymmetry ($\beta\rightarrow1$), however, this almost resembles the twin-field QKD configuration where GHz repetition rates can be used in principle. Also note that for a maximal repetition rate of 1 MHz the completely asymmetric scheme no longer outperforms the symmetric one.} 
\label{fig:asym}
\end{figure}

Notice that in this general case $f$ is no longer a real number and it is even possible that $\Re(f)=0$. Therefore, the secret-key fraction may become zero in this simple error model. As it can be seen in Fig. \ref{fig:asym}, for a fixed total distance, the secret-key rate oscillates with respect to $\beta$ including an envelope. The oscillations originate from the fact that $\Im(f)\neq0$ is possible. The envelope takes the following form: for $\beta$<$\beta_{max}$ it increases with $\beta_{max}>\frac{1}{2}$, while it drops when $\beta>\beta_{max}$. This comes from the gain in repetition rate while not losing too much from the worse scaling per channel use. By further increasing $\beta$ the envelope now decreases due to the worse scaling. In the region of $\beta\approx 1$ the secret-key rate per second rapidly increases again, because the repetition rate grows quickly up to the point where it is limited by the possible repetition rate of the light source. Due to the oscillations it is necessary to optimize $\beta$ for any given total distance. When considering increasing distances, $\beta_{max}$ moves nearer to $\frac{1}{2}$ and the advantage compared to the symmetric case of $\beta=\frac{1}{2}$ becomes less pronounced. For total distances of 200 km, we can increase the secret-key rate by 4.6\%, while for a total distance of 400 km we only gain 1.1\%.

\vspace{0.3cm}
\section{Calculation of the quantum repeater states with homodyne measurements}
\label{app:homodyne}
Let us first start with the no-loss case and again consider the state
\begin{align}
\frac{1}{2}(\ket{\uparrow,\uparrow,\alpha e^{-i\theta},\alpha e^{-i\theta}}+\ket{\downarrow,\downarrow,\alpha e^{i\theta},\alpha e^{i\theta}}\\ \nonumber
+\ket{\downarrow,\uparrow,\alpha e^{i\theta},\alpha e^{-i\theta}}+\ket{\uparrow,\downarrow,\alpha e^{-i\theta},\alpha e^{i\theta}}) \,.
\end{align}
After applying the beam splitter and the measurement of $\hat{p_1}=p$ and $\hat{x}_2=x$  we have the conditional two-qubit state (after tracing out the optical modes)
\begin{align}
\frac{1}{2}(\ket{\uparrow,\uparrow}\bra{\hat{p}=p}\ket{\sqrt{2}\alpha e^{-i\theta}}\bra{\hat{x}=x}\ket{0}+\\ \nonumber\ket{\downarrow,\downarrow}\bra{\hat{p}=p}\ket{\sqrt{2}\alpha e^{i\theta}}\bra{\hat{x}=x}\ket{0}+\\ \nonumber
\ket{\downarrow,\uparrow}\bra{\hat{p}=p}\ket{\sqrt{2}\alpha\cos{\theta} }\bra{\hat{x}=x}\ket{i \sqrt{2}\alpha\sin{\theta} }+\\ \nonumber\ket{\uparrow,\downarrow}\bra{\hat{p}=p}\ket{\sqrt{2}\alpha\cos{\theta} }\bra{\hat{x}=x}\ket{-i \sqrt{2}\alpha\sin{\theta} }) \,.
\end{align}
As the next step we calculate position- and momentum-space wave functions of a coherent state with amplitude $x_0+i p_0$. In order to express these wave functions in terms of vacuum-state wave functions of the harmonic oscillator we will make use of the displacement operator ($\hbar=\frac{1}{2}$ in our notation) and the Baker-Campbell-Hausdorff formula:
\begin{align} 
&\bra{\hat{x}=x}\ket{x_0+i p_0} \\ \nonumber&=\bra{\hat{x}=x}\exp((x_0+i p_0)(\hat{x}-i \hat{p})-(x_0-i p_0)(\hat{x}+i\hat{p}))\ket{0}\\ \nonumber
&=\bra{\hat{x}=x}\exp(2i (p_0 \hat{x}-x_0\hat{p}))\ket{0}\\ \nonumber
&=\bra{\hat{x}=x}\exp(2ip_0 \hat{x})\exp(-2ix_0\hat{p})\exp(-i p_0 x_0)\ket{0}\\ \nonumber
&=\bra{\hat{x}=x-x_0}\ket{0}\exp\left[2 i p_0(x-\frac{x_0}{2})\right]\\ \nonumber
&=\sqrt[4]{\frac{2}{\pi}} \exp\left[-(x-x_0)^2\right]\exp\left[2 i p_0(x-\frac{x_0}{2})\right]\,.
\end{align}
Similarly, one can show
\begin{widetext}
\begin{align}
\bra{\hat{p}=p}\ket{x_0+i p_0}=\sqrt[4]{\frac{2}{\pi}} \exp\left[-(p-p_0)^2\right]\exp\left[-2 i x_0(p-\frac{p_0}{2})\right].
\end{align}
\end{widetext}
We postselect onto states where $p\in(-\Delta_p,\Delta_p)$ and $x\in(-\Delta_x,\Delta_x)$. Further we label the density matrix elements in the same way as in the case with on/off detectors (see Appendix \ref{app:calc}) and we obtain the following results (all elements must be divided by the matrix trace, $2(a+b)$ , for normalization; for brevity we also omitted some extra factors which cancel anyway then through normalization),
\begin{widetext}
\begin{align}
a&=\frac{1}{2}\left(\text{erf}(\sqrt{2}\Delta_p-2\alpha \sin(\theta))+\text{erf}(\sqrt{2}\Delta_p+2\alpha \sin(\theta))\right) \,,\\
b&=\text{erf}(\sqrt{2}\Delta_p) \,, \\
c&=\exp\left(2\alpha^2(-1+\exp(2 i \theta))\right)\text{erf}(\sqrt{2}\Delta_p) \,, \\
f&=\exp\left(-4\alpha^2 \sin[2](\theta)\right)\text{erf}(\sqrt{2}\Delta_p)\frac{\text{Re}\left[\text{erf}(\sqrt{2}\Delta_x+2 i \alpha \sin(\theta)\right)]}{\text{erf}(\sqrt{2}\Delta_x)}\,.
\end{align} 
\end{widetext}
When including loss we can make use of Eq. (\ref{eq:tracetrick}) and after simplifications one can see that the expressions for $a,b,c,f$ almost stay the same. We only have to replace $\alpha\rightarrow \alpha \sqrt{\sqrt{\eta}}$ within the erf-functions and otherwise nothing changes where $\sqrt{\eta}$ is the transmission parameter corresponding to one physical segment (half a repeater segment). For example, for $n=1$, we have $\alpha\rightarrow \sqrt[4]{\eta_{total}}\alpha$. Using the expressions $a,b,c,f$ we can then calculate the BB84 secret-key fraction as before (we did not explicitly calculate $d_1$ and $d_2$, because we only need their values when considering $n>1$ and also not discarding the off-diagonal terms in the Bell basis).
\section{Different distribution/swapping strategies}
\label{app:sequential}

Let us discuss the effects of memory dephasing for the sequential and parallel entanglement distribution schemes. First of all we have to point out that the choice of $M_\text{par}$ is not optimal for more than two segments, because it assumes that the entanglement swapping operations are performed at the end, only after the entanglement distributions in all segments have succeeded. 
To illustrate this point, let us consider the example that first two adjacent segments succeeded and we have to wait one more time step until all the other segments succeeded  so that we can perform all swapping operations. This means the value of $M$ would be 4, because two segments (with two memories each) waited for one time step. Instead, we could also consider the case that we first perform the swapping operation on the two segments immediately after their successful creations and after the extra single time step we perform the remaining swapping operations. As a consequence, the value of $M$ is only 2, because only two memories waited for one time step.
This means it is beneficial to swap as soon as possible in order to keep the number of dephasing memories low 
\footnote{However, note that if we assumed probabilistic entanglement swapping instead of a deterministic one, swapping as soon as possible would yield a non-optimal raw rate, because one does not want to perform many entanglement swapping operations between entangled pairs of long and short distance since if the operation fails all involved segments have to start from scratch. }.

 Unfortunately, it is currently not even known how to calculate the probability distribution of $M=M_{par}$ for $n>2$ in the simple case where we wait for the success of all segments before performing the swapping operations.
If we want to consider more than two segments in a parallel distribution scheme, however, we can use the bound  $\mathds{E}[\exp(-M\frac{\tau}{T})]\geq \exp[-\mathds{E}(M)\frac{\tau}{T}]$ which can be obtained by applying Jensen's inequality.
As the expectation value operation is linear, we can easily calculate $\mathds{E}(M)$ since the exact $\mathds{E}(\max(X_1,\cdots,X_n))$ is already known in the literature \cite{hybridrepeaterrateanalysis}, and we obtain (for the case when Alice and Bob do not store their halves, so for $M$ from Eq. (\ref{eq:par_dephasing_short})):
 \begin{equation}
 \mathds{E}(M_{par})=2(n-1)\left(\sum_{j=1}^{n}\binom{n}{j}\frac{(-1)^{j+1}}{1-q^j}-\frac{1}{p}\right)\,,
 \end{equation}
 also using the well-known result for a geometrically distributed variable, $\mathds{E}(X_j)=\frac{1}{p}, \forall j=1\dots n.$
 We can use the inequality in order to obtain a lower bound on the secret-key fraction.
 However, one needs to bear in mind that this is only a lower bound that becomes very loose in the regime of bad memories.
 For the simple case of $n=2$, we calculated  $\exp[-\mathds{E}(M)\frac{\tau}{T}]$  and $\mathds{E}[\exp(-M\frac{\tau}{T})]$ (see App.\ref{app:dephasing}) and compared their corresponding secret-key fractions (assuming $p=10^{-4}, \sqrt{\eta}\ll1$).
 For the case of $T=10\mathds{E}(M)\tau$, we found that the exact calculation yields a 1\% higher secret-key rate. When considering $T=\mathds{E}(M)\tau$ the error increased to 86\% and when looking at memories with  $T=0.1\mathds{E}(M)\tau$ the approximation underestimated the secret-key fraction by six orders of magnitude, although the exact secret-key fraction of $2\times10^{-3}$ was not ridiculously low.
  Numerical simulations show that the bound becomes tighter for an increasing number of repeater segments. Unfortunately, realistic coherence times are often too small for obtaining a good bound by applying Jensen's inequality.
 
Let us now discuss the difference between a sequential and a parallel scheme with respect to the secret-key fraction. 
In order to be sure that improvements in the state quality arise from the changed strategy and not only from using the exact expression instead of a lower bound, we will now also compare the two strategies using for both the lower bound based on Jensen's inequality (for the sequential scheme, in addition, we use the exact rates).
For simplicity, let us consider the case where Alice and Bob perform the measurements on their qubits at the end after the entanglement was distributed over the whole distance and define the random variable $M_\text{seq} := 2\sum_{j=2}^n X_j$ (in the other case the sequential scheme also has a larger improvement than the parallel one). We then have
\begin{align}
\mathds{E}[M_\text{seq}]=2\frac{n-1}{p}\,,\\
\mathds{E}[M_\text{par}]\approx 2n\frac{H(n)-1}{p}\,,
 \end{align}
where $M_\text{par}$ is taken from Eq.(\ref{eq:par_dephasing_long}) and we used the approximation for the parallel scheme derived in App. \ref{app:approx} assuming $p\ll1$. For $n=2$ the protocols are the same and it can easily be checked that the sequential protocol is better for $n\geq3$. Better here means that less memory time is needed leading to a better secret-key fraction. Which protocol is the best in terms of the secret-key rate also depends on the memory coherence time $T$. If we have perfect memories ($T=\infty$), we do not gain any advantage due to the sequential protocol, but we have the disadvantage of a lower raw rate ($\frac{p}{n}$ versus $\frac{p}{H(n)}$), resulting in a lower overall secret-key rate. Note that for $n=2$ when we use the exact dephasing expressions for both the parallel and the sequential schemes, the parallel one even has a smaller dephasing than the sequential one, as already  pointed out in App. \ref{app:discuss}.

\begin{figure}[h]
\includegraphics[width=0.4\textwidth]{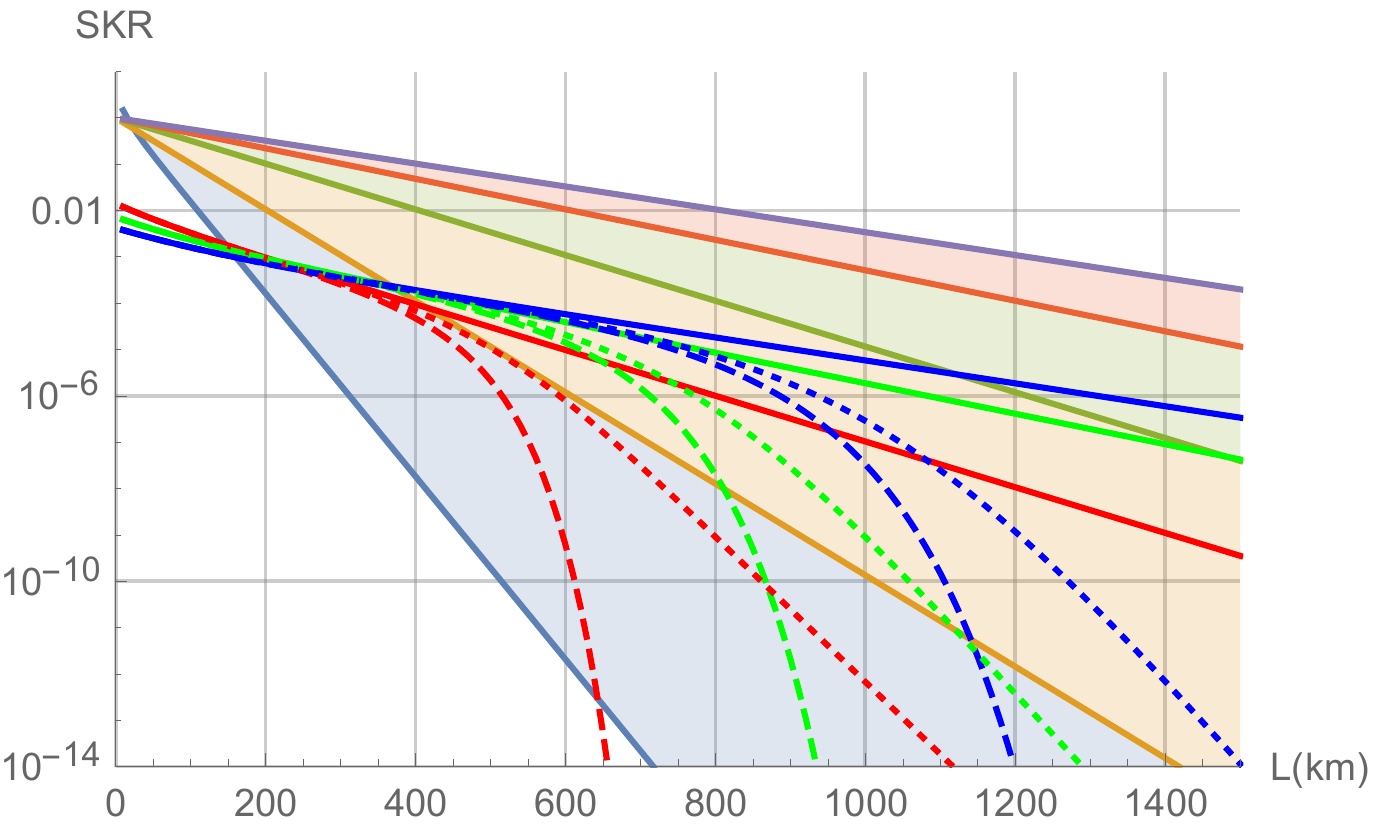}
\caption{(Color online) Secret-key rate for a repeater with $n$=2 (red), 3 (green), 4 (blue) (from left to right in terms of dropping secret-key rate)  segments using a sequential protocol ($\alpha=23.9$ in all cases). The undashed lines show the ideal loss-only case ($p_\text{det}=1$), while the dashed lines correspond to the case where we additionally consider a finite memory coherence time of 10 seconds. The dotted lines use the exact expression for the expectation value of the dephasing. The benchmarks (from bottom to top) PLOB, $\sqrt{\eta_\text{tot}}$,  $\sqrt[4]{\eta_\text{tot}}$,  $\sqrt[6]{\eta_\text{tot}}$ and  $\sqrt[8]{\eta_\text{tot}}$ can also be seen. The regions between two of those benchmarks are highlighted in color.}
\label{fig:jensen_seq}
\end{figure}

\begin{figure}[h]
	\includegraphics[width=0.4\textwidth]{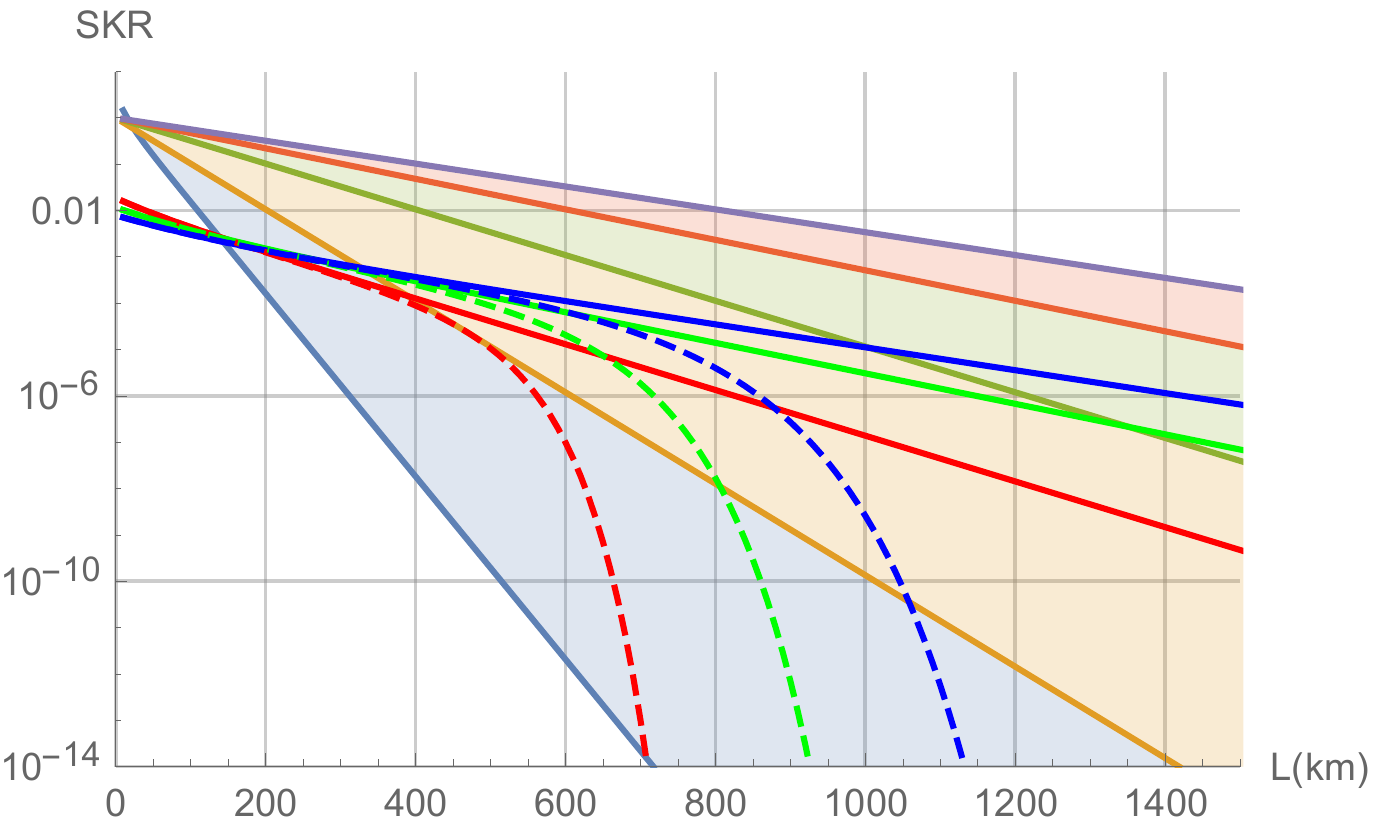}
	\caption{(Color online) Secret-key rate for a repeater with $n$=2 (red), 3 (green), 4 (blue) (from left to right in terms of dropping secret-key rate)  segments using a parallel protocol ($\alpha=23.9$ in all cases). The undashed lines show the ideal loss-only case ($p_\text{det}=1$), while the dashed lines correspond to the case where we additionally consider a finite memory coherence time of 10 seconds using Jensen's inequality. The benchmarks (from bottom to top) PLOB, $\sqrt{\eta_\text{tot}}$,  $\sqrt[4]{\eta_\text{tot}}$,  $\sqrt[6]{\eta_\text{tot}}$ and  $\sqrt[8]{\eta_\text{tot}}$ can also be seen. The regions between two of those benchmarks are highlighted in color.}
	\label{fig:jensen_par}
\end{figure}

 The obtainable secret-key rate using Jensen's inequality for the sequential and parallel protocols with a memory coherence time of 10s can be seen in Figs. \ref{fig:jensen_seq} and \ref{fig:jensen_par}. It can be seen that for $n=2$ the parallel scheme is superior, because both schemes have the same amount of dephasing but the parallel scheme has a better raw rate. However, for $n=3$ the rates of both schemes are quite similar and for $n=4$ the sequential scheme outperforms the parallel one as one might anticipate due to the better dephasing. Clearly, when using the exact expression for the dephasing in the sequential scheme we obtain significantly better rates than for the parallel scheme with rates calculated from the lower bound. However, for n>2, the rates of the sequential scheme based on the lower bound are still at least as good or even better ($n>$3) than those for the parallel scheme. This is our motivation for employing the sequential scheme throughout whenever we consider $n>2$ (besides the benefit that this allows us to compute the exact rates also for larger schemes, $n>2$).
\newpage
\bibliography{ref}
\end{document}